\documentclass[article,twocolumn,floatfix,superscriptaddress,longbibliography]{revtex4-2}
\usepackage{physics}
\usepackage[colorlinks,citecolor=blue,linkcolor=blue,urlcolor=blue]{hyperref}
\setcitestyle{super}
\usepackage{graphicx}
\usepackage{color}

\usepackage[colorlinks,citecolor=blue,linkcolor=blue,urlcolor=blue]{hyperref}
\usepackage{siunitx}
\usepackage{bbold}
\usepackage{indentfirst}
\usepackage{amsmath}
\usepackage{amsmath,amssymb,amsfonts}
\usepackage{mathrsfs}%
\newcommand{\Nel}{N_{\mathrm{el}}}

\newcommand{\m}{\mathrm{m}}
\usepackage{dutchcal}

\begin{document}
%\title{Enhanced fractional quantum Hall gaps and exchange splitting reduction in a 2D electron gas via a hovering split-ring resonator}
\title{Enhanced fractional quantum Hall gaps in a two-dimensional electron gas  \\ 
coupled to a hovering split-ring}
\author{Josefine Enkner}
\affiliation{Institute for Quantum Electronics, ETH Zurich, CH-8093 Zurich, Switzerland}
\affiliation{Quantum Center, ETH Zürich, 8093 Zürich, Switzerland}

\author{Lorenzo Graziotto}
\affiliation{Institute for Quantum Electronics, ETH Zurich, CH-8093 Zurich, Switzerland}
\affiliation{Quantum Center, ETH Zürich, 8093 Zürich, Switzerland}

\author{Dalin Boriçi}
\affiliation{Université Paris Cité, CNRS, Matériaux et Phénomènes Quantiques, 75013, Paris, France}

\author{Felice Appugliese}
\affiliation{Institute for Quantum Electronics, ETH Zurich, CH-8093 Zurich, Switzerland}
\affiliation{Quantum Center, ETH Zürich, 8093 Zürich, Switzerland}

\author{Christian Reichl}
\affiliation{Laboratory for Solid State Physics, ETH Zürich, 8093 Zürich, Switzerland}

\author{Giacomo Scalari}
\affiliation{Institute for Quantum Electronics, ETH Zurich, CH-8093 Zurich, Switzerland}
\affiliation{Quantum Center, ETH Zürich, 8093 Zürich, Switzerland}

\author{Nicolas Regnault}
\affiliation{Laboratoire de Physique de l’Ecole normale sup\'erieure,ENS, Universit\'e PSL, CNRS, Sorbonne Universit\'e}
\affiliation{Department of Physics, Princeton University, Princeton, NJ 08544, USA}

\author{Werner Wegscheider}
\affiliation{Laboratory for Solid State Physics, ETH Zürich, 8093 Zürich, Switzerland} 

\author{Cristiano Ciuti}
\affiliation{Université Paris Cité, CNRS, Matériaux et Phénomènes Quantiques, 75013, Paris, France}

\author{Jérôme Faist}
\affiliation{Institute for Quantum Electronics, ETH Zurich, CH-8093 Zurich, Switzerland}
\affiliation{Quantum Center, ETH Zürich, 8093 Zürich, Switzerland}

%\begin{abstract}{ \noindent We present experimental findings on quantum Hall transport in a high-mobility two-dimensional electron gas coupled to a hovering split-ring resonator, without external illumination. Our investigations reveal a striking enhancement of fractional quantum Hall energy gaps alongside a concurrent reduction in exchange splitting at odd integer filling factors. We attribute these effects to the emergence of an effective cavity-mediated electron-electron long-range attractive interaction mediated by the exchange of virtual photons in the presence of significant spatial gradients of the cavity modes. These results unveil a compelling interplay between cavity quantum electrodynamics and electronic correlations in two-dimensional systems, with profound implications for the manipulation and control of quantum phases 2D materials.} \end{abstract}

\begin{abstract}{ \noindent The magnetotransport of a high-mobility two-dimensional electron gas coupled to a hovering split-ring resonator with controllable distance is studied in the quantum Hall regime. The measurements reveal an enhancement by more than a factor 2 of the quantum Hall energy gaps at the fractional filling factors 4/3, 5/3, and 7/5, alongside a concurrent reduction in exchange splitting at odd integer filling factors. 
Theoretically, we show the strength of both effects to be quantitatively compatible with the emergence of an effective electron-electron long-range attractive interaction mediated by the exchange of virtual cavity photons in the presence of significant spatial gradients of the cavity electric vacuum fields. 
%We attribute these effects to the emergence of an effective electron-electron long-range attractive interaction mediated by the exchange of virtual cavity photons in the presence of significant spatial gradients of the cavity electric vacuum fields. 
These results unveil a compelling interplay between cavity quantum electrodynamics and electronic correlations in two-dimensional systems, with profound implications for the manipulation and control of quantum phases in 2D materials.}
\end{abstract}

\date{\today}
\maketitle
In quantum mechanics, empty space is not devoid but is characterized by \emph{vacuum field fluctuations}, which underlie phenomena such as the Lamb shift~\cite{lambShift}, spontaneous emission, and the Casimir effect~\cite{milonni1994quantum}. Traditionally overlooked in condensed matter physics due to their quantitatively small relative contributions in free space atomic physics, the scenario changes dramatically when collective excitations are considered inside electromagnetic cavities where enhanced vacuum fields significantly boost electron-photon interactions. This interaction can alter the properties of solid-state matter and may induce novel material phases. Leveraging the ultra-strong light-matter coupling regime~\cite{ciuti2005quantum}, researchers aim to modify molecular structures~\cite{cavityMoleculeStrong}, enhance electron-phonon couplings~\cite{EPCcavity}, and facilitate the emergence of new electronic phases, including superconductivity~\cite{schlawin2019cavity}, ferroelectricity~\cite{cavityFerroelectric}, and topological properties~\cite{nguyen2023electron,winter_fractional}. Despite substantial theoretical backing and evidence of cavity-induced alterations in chemical reactions~\cite{ribeiro2018polariton, ahn2023modification} and charge transport~\cite{cavityTransport, paravicini2019magneto,ciuti2021cavity, appugliese2022breakdown,winter_fractional}, demonstration of the manipulation of strongly correlated phases~\cite{Girvin_2019} by cavity vacuum fields remain scarce.

\begin{figure*}[t!]
\includegraphics[width=1\textwidth]{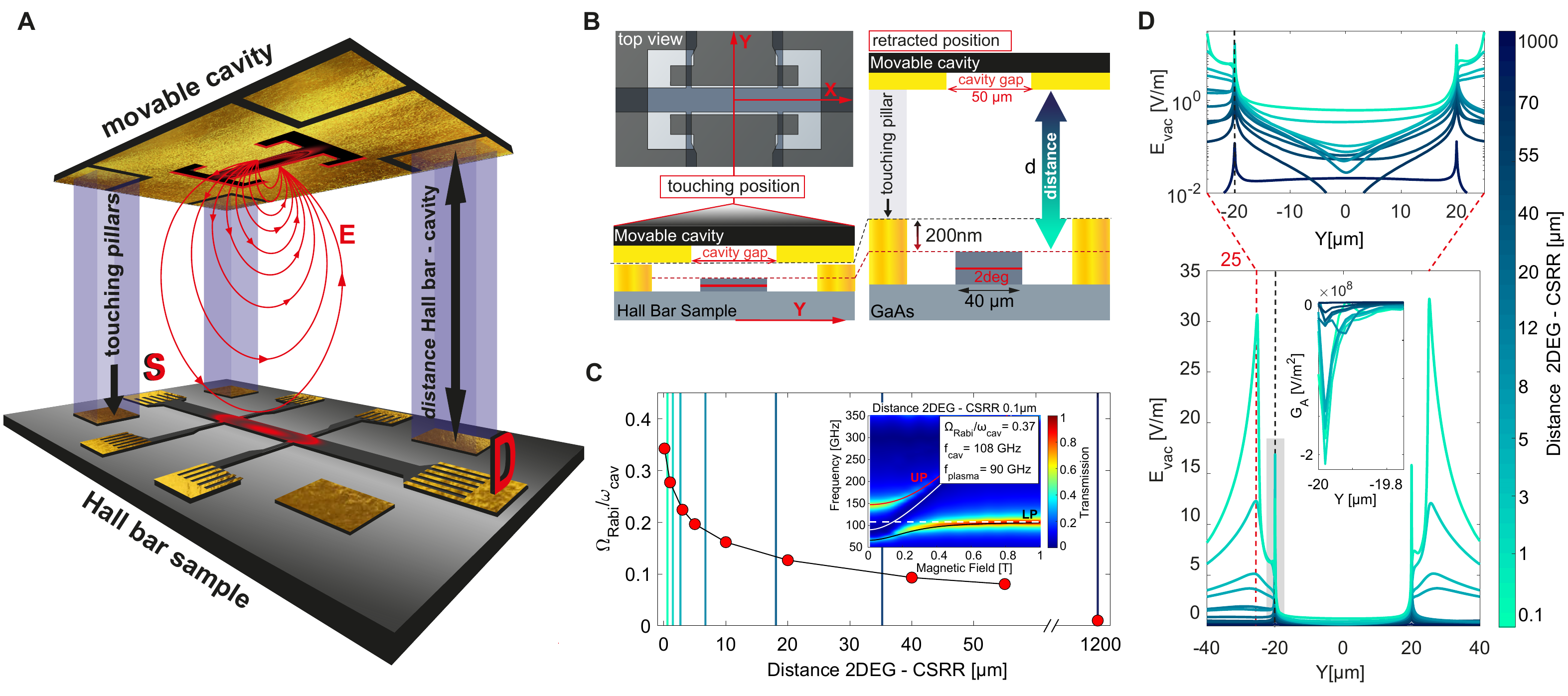}
\caption{\textbf{Description of the experimental setup consisting of a movable split-ring resonator hovering over a Hall bar}. \textbf{(A)} Artistic rendering of the platform implemented to tune the coupling strength between the 2D electrons within a Hall bar and the vacuum electromagnetic fields of a complementary split-ring resonator (CSRR). The CSRR is defined by a cutout in a gold metallic layer and approaches the Hall bar from above, therefore increasing the coupling to its fringing electric field indicated via the red arrows. \textbf{(B)} Top-left: Overlayed view of the Hall bar and CSRR (gray shading corresponds to the metal of the CSRR), which is achieved after aligning the markers on the back of the resonator plane with the Hall bar on the sample. Bottom and right: side profile in $y$ direction of the setup when the resonator plane and the conductive pillars protruding from the Hall bar sample are in close contact (touching position) and when the resonator plane is distant (retracted position). \textbf{(C)} Simulation of the normalized coupling strength over the distance between the resonator plane and the 2D electron gas. The colored vertical lines indicate the distances $d$ for which the transport traces in Fig.~\ref{fig_FQH_gaps}A and Fig.~\ref{fig_spin}A were measured in the experiment. Inset: simulated polariton dispersion of the coupled system with the cavity being $0.1~\si{\micro\m}$ away from the 2DEG. \textbf{(D)} Simulation of the cavity vacuum electric field profile across the Hall bar at B=0.3~T. The red dashed line indicates the gap region of the CSRR, and the black dashed line marks the edge of the Hall bar. Top: zoom in on the area around the Hall bar. Inset: Gradient of the vacuum electric field exponentially decaying into the 2DEG (shaded area).}
\label{fig1}
\end{figure*}

In this work, as shown in Fig. \ref{fig1}, we have developed a mobile cavity system capable of finely tuning cavity vacuum fields through the controlled adjustment of the metallic split-ring resonator's distance from a high-mobility GaAs-based two-dimensional (2D) electron gas within a quantum well heterostructure. When subjected to a strong perpendicular magnetic field, this setup allows for the probing of strongly correlated fractional quantum Hall phases via transport measurements of longitudinal and transverse resistance. Remarkably, we have discovered that the ``hovering" cavity resonator significantly enhances several fractional quantum Hall phases, effectively more than doubling their corresponding many-body energy gaps. This enhancement occurs while maintaining the metallic resonator at distances where electrostatic screening effects are entirely negligible.

The energy gaps of fractional quantum Hall states are known to be sensitive to the form of electron-electron interactions. Previous theoretical research \cite{Abanin2011} has highlighted the potential to enhance these gaps in mono-layer 2D materials such as graphene. This enhancement typically leverages the electrostatic screening effects of a closely positioned dielectric layer—on the order of the electron's magnetic length—to enhance fractional quantum Hall gaps. However, such a strategy is generally infeasible for GaAs quantum wells, which are embedded within a thicker semiconductor layer that prevents close proximity adjustments. Furthermore, the substitution of a dielectric layer with a metallic one is expected to reduce the gap \cite{Abanin2011}$^{,}$ \footnote{We have independently recalculated the fractional quantum Hall gaps taking into account the Coulomb potential modified by the image charges produced by a metallic plate and found agreement with Ref. \cite{Abanin2011} for their case $\alpha = -1$. In this case, there is a reduction of the Laughlin gap. Moreover, when the distance is much larger than the magnetic length, like in our experiments, the magnitude of the change is totally negligible. Additional details can be read in the Supplementary Material.}. 

In our experiment, we utilize a metallic resonator positioned at a distance from the 2D electron gas, typically orders of magnitude larger than the magnetic length, a range at which electrostatic modifications to the Coulomb potential become negligible. Moreover, contrary to expectations of electrostatics in the presence of a metal plate, the metallic resonator does not decrease but rather increases the fractional quantum Hall gaps. We interpret this remarkable effect as stemming from the emergence of a cavity-induced long-range attractive electron-electron potential mediated by the exchange of virtual cavity photons. In particular, the presence of strong spatial gradients of the vacuum fields is essential to create cavity-mediated electron-electron interaction within the same Landau level.

%theoretical suggestion that the probability for an electron to scatter via the anti-resonant process depends on the distance between the Fermi energy and the center of the band which is smaller for the case when the gap between the $n$-th LL and the adjacent state corresponds to the Zeeman gap.
%In the experiment, we observed that especially the odd filling factors were affected by cavity-mediated hopping, which we attributed to the theoretical suggestion that the probability for an electron to scatter via the anti-resonant process depends on the distance between the Fermi energy and the center of the band which is smaller for the case when the gap between the $n$-th LL and the adjacent state corresponds to the Zeeman gap. %Instead, this work now shows that the cavity, in fact, is directly decreasing the Zeeman gap. 

%Part of our experimental findings support the suggestion that cavity-induced electric field gradients $G_A$ localized at the edge of the Hall bar play a leading role in permitting back-scattering processes through cavity-induced "optical disorder"~\cite{ciuti2021cavity} and therefore implies that the effect is localized at the edges.
\begin{figure*}[t!]
\centering
\includegraphics[width=1\textwidth]{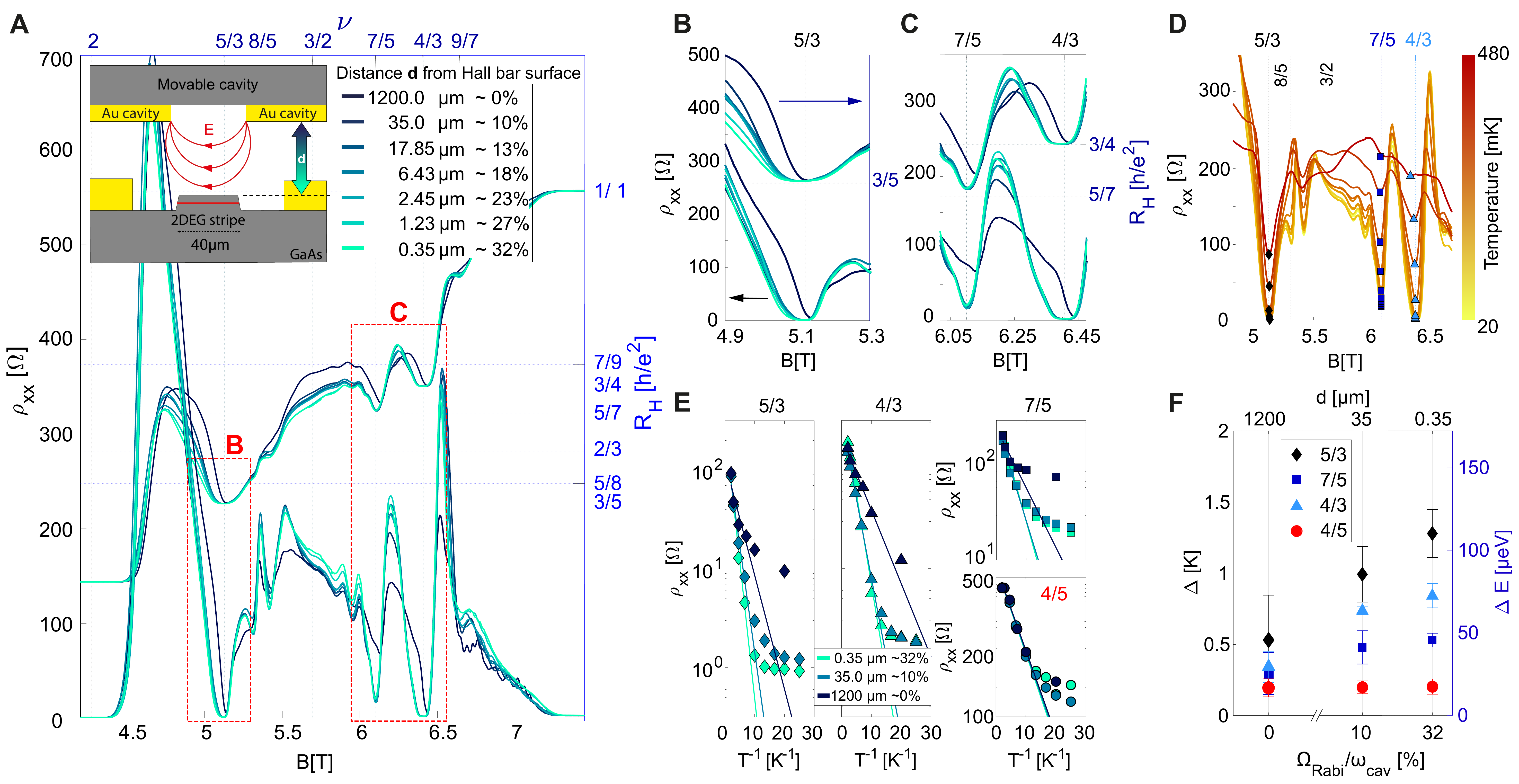}
\caption{\textbf{Cavity-enhanced fractional quantum Hall energy gaps}  \textbf{(A)} Longitudinal resistance (left vertical scale) and Hall resistance (right vertical scale) in the magnetic field region exhibiting the fractional quantum Hall states. Inset: Side view scheme of the setup. \textbf{(BC)} Zoom in to the fraction 5/3, 7/5, and 4/3.
Longitudinal magnetoresistance versus magnetic field B at low magnetic fields for different distances between the movable cavity and the Hall bar. \textbf{(D)} Longitudinal resistance versus magnetic field for different temperatures with the hovering cavity being $0.35~\si{\micro\meter}$ away from the Hall bar. \textbf{(E)} Arrhenius activation plots for different fractional filling factors for the three distances $0.35~\si{\micro\meter}$, $35~\si{\micro\meter}$ and $1200~\si{\micro\meter}$ with corresponding coupling $32\%$, $10\%$, and $\approx0\%$. \textbf{(F)} Extracted activation energy gaps for fractional fillings $5/3$, $7/5$, and $4/3$ as a function of the distance between the split-ring resonator and the Hall bar.}
\label{fig_FQH_gaps}
\end{figure*}

The experimental setup depicted in Fig.~\ref{fig1}A is designed to vary the spacing between a Hall bar and a cavity, specifically a Complementary Split-Ring Resonator (CSRR) evaporated on a GaAs substrate. Adjusting this distance allows us to modulate the intensity of the cavity's fringing fields, which extend beyond the resonator's gap (illustrated by red lines). These fields penetrate the Hall bar and interact with the electron gas. Fig.~\ref{fig1}B shows the alignment of the CSRR's \SI{50}{\micro m} gap with the \SI{40}{\micro m} wide Hall bar, which was crafted using conventional photolithography on a GaAs-based heterostructure. The quantum well beneath the surface, located at a depth of $233~\si{\nano\meter}$, boasts a high electron mobility of $1.69\times 10^7\mathrm{cm^2 V^{-1}s^{-1}}$ and a sheet density of $2.06\times 10^{11}\mathrm{cm^{-2}}$ at 1.3K under dark conditions.

Adjacent to the Hall bar are four etched pillars rising $200~\si{\nano\meter}$ above the surface, effectively spacing the CSRR's gold plane from the Hall bar to prevent physical contact and serving as alignment references for the initial setup. Simulations, detailed in Fig.~\ref{fig1}C, explore the coupling strength $\Omega_\mathrm{Rabi}/\omega_{\mathrm{cav}}$—where $\Omega_\mathrm{Rabi}$ represents the Rabi frequency~\cite{hagenmuller2010ultrastrong} and $\omega_{\mathrm{cav}}$ denotes the cavity's angular frequency. By fitting the Hopfield model's anti-crossing curve to the polaritonic dispersion (inset of Fig.1C), it is demonstrated that the system can be finely adjusted from a near decoupled state ($\Omega_\mathrm{Rabi}/\omega_{\mathrm{cav}}\approx0$ at a $1200~\si{\micro\meter}$ cavity distance) to an ultra-strong coupling regime\cite{scalari2012ultrastrong} ($\Omega_\mathrm{Rabi}/\omega_{\mathrm{cav}}=37 \%$ as the cavity approaches within $0.1\si{\micro\meter}$ of the 2DEG). As the resonator approaches, the cavity's resonance frequency shifts downward from $145~\si{\giga\hertz}$ to $115~\si{\giga\hertz}$, 
%a change attributed to capacitive coupling and alterations in the effective dielectric constant. 
a change attributed to the fact that the mode field reaches the semiconductor region.
Note that measurements were taken at a minimal CSRR-to-Hall bar surface distance of $d=0.35~\si{\micro\meter}$, achieving a coupling strength of $31 \%$, where $d$ indicates the separation between the CSRR and the Hall bar surface, ensuring the resonator plane and the Hall bar sample remain spatially separate (Fig.\ref{fig1}B). In contrast, the simulations pertain to the CSRR's distance from the 2DEG, which is effectively increased by an additional $250\,\si{\nano\meter}$ semiconductor layer atop the 2D electron gas. 

%, one should add $~250~\si{\nano\meter}$.
%We remark that in the experiment, the minimum distance between the CSRR and the 2DEG is defined by the pillars and depth of the 2DEG within the heterostructure, so $\approx 0.45~\si{\micro\meter}$, whereas measurements were performed for minimum distance $d=0.35~\si{\micro\meter}$, corresponding to a coupling of $0.31\%$, where $d$ is the distance between the CSRR and the surface of the Hall bar, such that the resonator plane and the Hall bar sample are spatially disconnected (Fig.~\ref{fig1}B).
%performed measurements where the resonator plane and the Hall bar sample are spatially disconnected, with the closest distance being $0.35~\si{\micro\meter}$ (Fig.~\ref{fig1}C).
%In the following, we display the experimental data as a function of the distance between the surface of the Hall bar and the resonator plane. We point out that in the position $0.35~\si{\micro\meter}$, the CSRR is not in contact with the pillars and, therefore, remains entirely spatially separate from the Hall bar (Fig.~\ref{fig1}C).
%As the resonator approaches, the resonance frequency of the cavity drops from $145~\si{\giga\hertz}$ when positioned $100~\si{\micro\meter}$ away to $115~\si{\giga\hertz}$ when close, due to changes in the effective dielectric constant. 
In Fig.~\ref{fig1}D, we study the vacuum electric field (\(E_{\text{vac}}\)) distribution across a Hall bar at a magnetic field of \(B=0.3~\text{T}\) for different distances, focusing on the lower polariton branch, while noting that the upper polariton branch shows very similar results as shown in the supplementary~\cite{supplementary}. 
%The vacuum field strength $E_\mathrm{vac}$ is obtained by normalizing the spatial field profile from the simulation to the vacuum field strength associated to the estimated Rabi frequency, using the formula \(\mathcal{E}_\mathrm{vac}=\frac{\hbar\Omega_\mathrm{Rabi}}{d\sqrt{N_e}}\)~\cite{hagenmuller2010ultrastrong} where $d$ is the dipole moment and $N_e$ the number of coupled electrons.
As the resonator plane is moved closer to the Hall bar, the amplitude of the electric field gradually increases. Around the Hall bar, centered at zero and spanning \SI{40}{\micro m}, four symmetric maxima emerge at the 2DEG level. Peaks at $\pm \SI{25}{\micro\meter}$ align with the CSRR resonator gap edges, while maxima at $\pm \SI{20}{\micro\meter}$ correspond to the Hall bar edges, where the CSRR’s vacuum electric field 
$E_{\mathrm{vac}}$ interacts with 2DEG's electrons and then rapidly decays into the bulk, leading to a maximum field gradient of the order of $10^8\mathrm{V/m^2}$ (see Fig.~\ref{fig1}D inset) and to a remaining electric field of the order of $0.8~\mathrm{V/m}$  (See Fig.~\ref{fig1}D) within the bulk of the Hall bar when the system is $37\%$ coupled. 

During a single cooldown, we measure the longitudinal resistivity $\rho_{xx}$ on the side of the Hall bar and the transverse resistances $R_H$ across the Hall bar as a function of the magnetic field. Each trace is taken for a specified distance $d$ and can be associated with the corresponding coupling according to the simulations in Fig.~\ref{fig1}C. Measurements are performed at an electronic temperature of $\approx20~\si{\milli\kelvin}$.

%As the resonator plane is moved away from the Hall bar, the amplitude of peaks at the edge of the CSRR and the Hall bar, the effective vacuum field in the bulk, and the magnitude of the field gradient diminishes correspondingly.

%\begin{figure}[t!]
%\centering
%\includegraphics[width=0.5\textwidth]{Figures_main/g3.pdf}
%\caption{\textbf{Cavity-induced reduction of the exchange splitting and effective g-factor} \textbf{(A)}. The plot of the activation energies extracted from the Arrhenius plot (inset) for odd filling factor $\nu=25$ to $\nu=13$. The effective g-factor is obtained from a linear fit indicated with dashed lines. Inset: Arrhenius plot and magnetic field range for filling factor $nu=15$ with the resonator plane being $0.35\si{\micro\meter}$ away from the surface of the Hall bar. \textbf{(B)} Evolution of the activation energy of $\nu=13$ as a function of the distance.}
%\label{fig3}
%\end{figure}
\begin{figure*}[t!]
\centering
\includegraphics[width=1\textwidth]{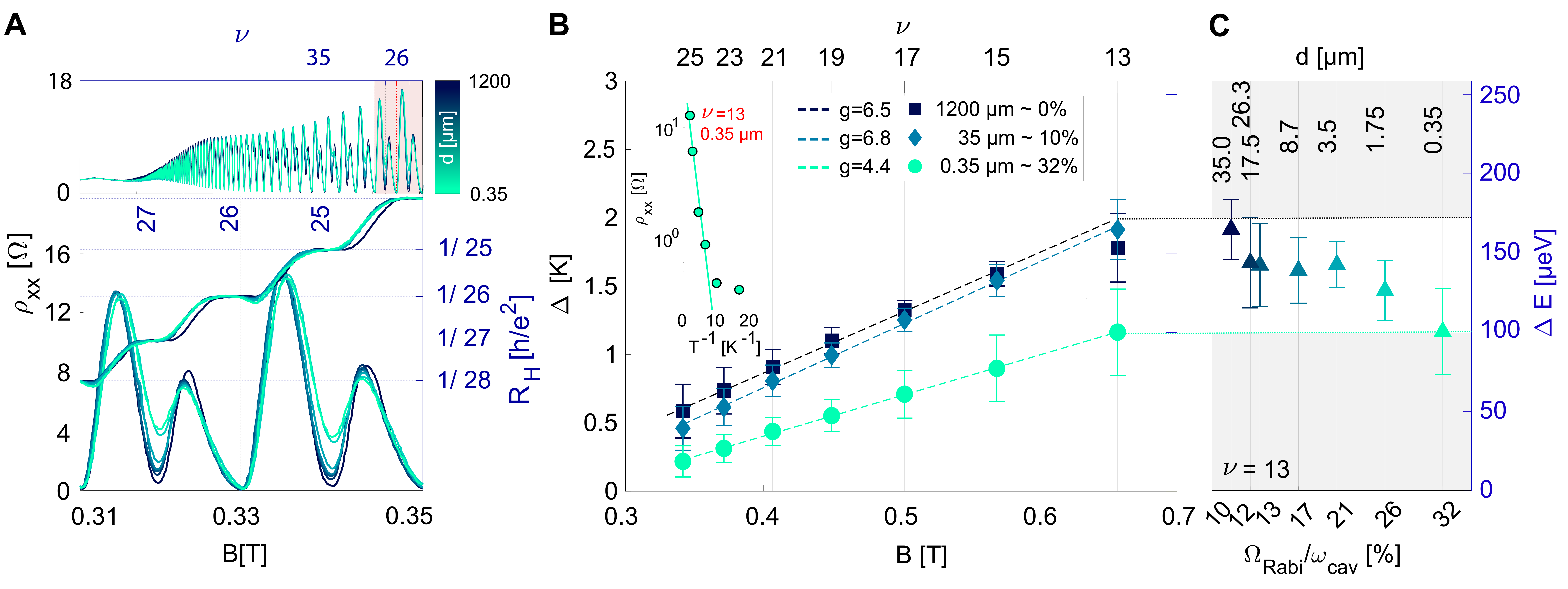}
\caption{\textbf{In situ modification of quantum Hall transport and cavity-induced reduction of the exchange splitting and effective g-factor.} \textbf{(A)} Longitudinal resistance (left vertical scale) and Hall resistance (right vertical scale) for different distances $d$. As the coupling increases, at odd integer values of the filling factor $\nu$, the longitudinal resistance at the minima rises while the correspondent quantum Hall plateaus lose quantization. \textbf{(B)} The plot of the activation energies extracted from the Arrhenius plot (inset) for odd filling factor $\nu=25$ to $\nu=13$. The effective g-factor is obtained from a linear fit indicated with dashed lines. Inset: Arrhenius plot for filling factor $\nu=13$ with the resonator plane being $0.35\si{\micro\meter}$ away from the surface of the Hall bar.  \textbf{(C)} Evolution of the activation energy of $\nu=13$ as a function of the coupling and corresponding distance $d$.} 
\label{fig_spin}
\end{figure*}

%% Results
%The Hall bar employed in our experiment has a width of \SI{40}{\micro m} and is fabricated using standard photolithography techniques on a GaAs-based heterostructure. The quantum well is buried $233~\si{\nano\meter}$ below the surface and exhibits a high electron mobility $1.69\times 10^7\mathrm{cm^2 V^{-1}s^{-1}}$ and a sheet density $2.06\times 10^{11}\mathrm{cm^{-2}}$ at 1.3K in dark conditions. The resonator plane is a CSRR fabricated on a GaAs substrate hovering above the Hall bar and can be tuned in its vertical distance to the Hall bar using precision nanopositioners (Attocubes).

The influence of the cavity vacuum fields on fractional quantum Hall states is profound. Remarkably, even at base temperature, we observe the impact of the cavity on the fractional states 5/3, 7/5, and 4/3, as depicted in Fig.~\ref{fig_FQH_gaps}ABC. With progressively increased coupling, we observe a noticeable widening of the plateaus alongside a reduction in the resistance minima of $\rho_{xx}$. We assess the energy gap by measuring the thermally activated longitudinal resistance (see Fig.~\ref{fig_FQH_gaps}D), which displays a characteristic exponential decline versus the inverse temperature depicted in the Arrhenius plots (Fig.~\ref{fig_FQH_gaps}E). As the distance is reduced from $1.2$ mm to $0.35~\si{\micro\meter}$, there is a noticeable increase in the energy gap for the fractional filling factors $5/3$, $7/5$, and $4/3$ within the second Landau level (see Fig.~\ref{fig_FQH_gaps}F). It is important to note that these fractions are part of the Jain principal sequence \cite{Jain1990} originating from the $1/3$ Laughlin state. Although our highest magnetic field strength does not permit us to reach the filling factor $\nu = 2/3$, we are able to achieve the fraction $4/5$, which is associated with the $1/5$ Laughlin state. Intriguingly, the energy gap for the $4/5$ fraction, observed in the lowest Landau level, does not increase significantly within the error bars. This suggests that the cavity-mediated effects may have contrasting impacts on the $1/3$ and $1/5$ families of fractional quantum Hall states.

Since fractional quantum Hall states arise solely from electron-electron interactions, these observations suggest that the cavity mediates an additional effective electron-electron interaction that competes with the Coulomb interaction. Such cavity-mediated interactions are also discernible at odd integer filling factors, where the activation energy gap is typically defined by spin splitting. In GaAs, this energy gap is predominantly influenced by the exchange splitting due to Coulomb interactions. The modifications in exchange splitting observed in our experiments indicate the presence of a cavity-mediated potential.

As illustrated in Fig.\ref{fig_spin}A top, at lower magnetic fields 
%the uniformity of density and mobility across all traces confirms the high quality of the sample and ensures that measurements are consistent across variables such as temperature and current passed through the Hall bar. 
and as the resonator plane is gradually brought closer to the Hall bar, increasing the coupling strength, we observe a consistent increase in the minima of the longitudinal resistivity and a gradual elevation of the plateaus (Fig.\ref{fig_spin}A bottom). 
%We observe the enhanced scattering rate on the even plateau by fitting the transverse resistance with a third-order polynomial function, showing the same exponential decay as a function of cyclotron energy as reported in \cite{appugliese2022breakdown}. 
%Initially, the resistivity at filling factor $\nu=21$ shows a zero-resistance state; however, this state is progressively diminished as the coupling intensifies (Fig.\ref{fig_spin}A, right). 
This behavior at odd filling factors suggests a reduction in effective spin splitting with increased cavity coupling. This effective spin splitting $\Delta$ is experimentally obtained by studying the thermally activated minima of the longitudinal resistance, which display a characteristic exponential decay as a function of the inverse temperature (see the Arrhenius plot in the inset of Fig.~\ref{fig_spin} B). As shown in Fig.~\ref{fig_spin}, we observe a linear dependence of the gap $\Delta$ as a function of the magnetic field whose slope $\mu_B g^{\star} / k_B$ (where $k_B  $ is Boltzman's constant and $\mu_B$ the Bohr magneton) directly yields~\cite{leadley1998critical} the effective gyromagnetic factor of the electron $g^{\star}$ whose value includes corrections from the exchange energy~\cite{janak1969g}.

%Although less pronounced, similar effects on the resistivity minima and plateau flatness are observed at even filling factors. We fit the transverse resistance around the plateaus with a third-order polynomial function and find that the slope of the fit shows the same exponential decay as a function of cyclotron energy~\cite{supplementary} as reported in \cite{appugliese2022breakdown}.

 We find the effective g-factor to decrease from $g^{\star}=6.5$ when the system is uncoupled (with the resonator plane about $\approx1.2$ mm away, Fig.~\ref{fig_spin}AB) to $g^{\star}=4.5$ when the system is ultra-strongly coupled, with the resonator plane positioned $\SI{0.35}{\micro\meter}$ from the Hall bar surface (approximately $\SI{0.6}{\micro\meter}$ from the 2DEG). In contrast with the case of the fractional quantum Hall regime, this experiment suggests that for odd quantum Hall plateau the modification of the electron-electron potential by the cavity has the effect of weakening the quantization, as was already observed in a more dramatic fashion in our previous experiments   \cite{appugliese2022breakdown} where the cavity was directly fabricated onto the Hall bar (see also the supplementary~\cite{supplementary} for further comparison between the two results). As recently reported~\cite{EnknerPRX}, we do not see evidence for a renormalization of the von Klitzing constant $R_k$ of either integer or fractional plateaus.   

%We assess the energy gap by measuring the thermally activated longitudinal resistance, which displays a characteristic exponential decline versus the inverse temperature, depicted in the Arrhenius plot (Fig.~\ref{fig_spin}A inset).

%We find that the energy gap at odd filling factors, represented by the Zeeman energy $E_\mathrm{Zeeman}=g\mu B+\Gamma$~\cite{leadley1998critical} accounts for Landau level (LL) broadening, $\mu$ is the Bohr magneton, and $B$ is the magnetic flux density—is markedly dependent on the coupling. This gap is quantified with an effective g-factor $g^{\star}$ that includes corrections from exchange energy. The effective g-factor decreases from $g^{\star}=6.5$ when the system is uncoupled (with the resonator plane about $\approx1.2$ mm away, Figure~\ref{fig_spin}AB) to $g^{\star}=4.5$ when the system is ultra-strongly coupled, with the resonator plane positioned $\SI{0.35}{\micro\meter}$ from the Hall bar surface (approximately $\SI{0.6}{\micro\meter}$ from the 2DEG). Complementing data for an additional Hall bar sample coupled to another CSRR can be found in the supplementary~\cite{supplementary}.
%For an additional Hall bar on the sample chip coupled to a different resonator with a resonance frequency of $\omega_\mathrm{slot}=200~\si{\giga\hertz}$, we find the effective g factor, initially $g^{\star}=7.05$ as the cavity is fully retracted, to be lowered to $g^{\star}=4.2$ within close proximity of the cavity. 

To understand the remarkable cavity-mediated quantum effects, we present here a quantum field theoretical description. Let us consider $N_{\mathrm{el}}$ electrons subject to a magnetic field 
perpendicular to the plane. For simplicity, we consider a model with a single cavity mode.
Including the Coulomb electron-electron repulsive potential 
$\hat{v}_{ij}$, the corresponding cavity quantum electrodynamical (QED) Hamiltonian reads:
\begin{equation}
    \hat{H} = 
    \hbar \omega_{\mathrm{cav}} \hat{a}^{\dagger}\hat{a} 
    +
    \sum_{i<j} \hat{v}_{ij}
    % \frac{1}{2} \sum_{i=1}^{N_{\mathrm{el}}} \sum_{j=1}^{N_{\mathrm{el}}} \hat{v}_{ij}
    +
    \frac{1}{2 m_{\star}} 
    \sum_{i = 1}^{\Nel}
    \left(
        \hat{\vb{p}}_i + e \hat{\vb{A}}_i
    \right)^2 \, .
\end{equation}
The operator $\hat{\vb{p}}_i$ 
is the momentum of the the $i$'th electron, 
$m_{\star}$ the effective electron mass and $-e$ its charge. 
The electromagnetic vector potential entering the Hamiltonian 
is the sum of two contributions.
One is associated with the static magnetic field $B$ and
one corresponds to the cavity mode, 
whose frequency is $\omega_{\mathrm{cav}}$ 
and whose bosonic creation (annihilation) operator 
is $\hat{a}^{\dagger}$ ( $\hat{a}$), namely
$
    \hat{\vb{A}}_i 
    =    
    \vb{A_0}(\hat{\bf{r}}_i) 
    +
    % \vb{A^{\mathrm{cav}}}(\hat{\bf{r}}_i)
    \hat{\vb{A}}^{\mathrm{cav}}(\hat{\bf{r}}_i)
    %A_{} (\hat{a} + \hat{a}^{\dagger}) \vb{e}_x 
    \, .
$
The part responsible for the magnetic field 
takes the following form in the symmetric gauge:
$
    \vb{A_0} (\hat{\bf{r}}_i) =  \frac{B}{2}(\hat{x}_i \vb{e}_y -\hat{y}_i \vb{e}_x ) \, ,
$
where $\hat{x}_i$ are $\hat{y}_i$ are the position operators for the $i$'th electron 
and $B$ is the amplitude of the applied static magnetic field $\vb{B} = B \vb{e}_z$.
For the cavity mode part, 
we assume for simplicity a field polarized along the $\vb{e}_x$ direction with a constant gradient 
along the $\vb{e}_x$ direction: 
$
    \hat{\vb{A}}^{\mathrm{cav}}(\hat{\bf{r}}_i)
    =
    \left(
        A_{\mathrm{vac}} + 
        \mathcal{G}_{\mathrm{A}} \hat{x}_i 
        % \mathcal{G}^{\mathrm{A}}_y \hat{y}_i
    \right)
    (
        \hat{a} + \hat{a}^{\dagger}
    ) 
    \vb{e}_x \, .
$
The gradient $\mathcal{G}_{\mathrm{A}}$
is related to the gradient of the electric field vacuum field through the relation
% $\mathcal{G}^{\mathrm{E}}_{x,y} =  \omega_{\mathrm{cav}} \mathcal{G}^{\mathrm{A}}_{x,y}$.
$\mathcal{G}_{\mathrm{E}} =  \omega_{\mathrm{cav}} \mathcal{G}_{\mathrm{A}}$.

The quantum light-matter interaction consists of a paramagnetic contribution (linear in the photon operators) and a diamagnetic term (quadratic in the bosonic operators).
The part of the Hamiltonian, which depends on the photonic operators only, can be diagonalized through a Bogoliubov transformation in terms of the renormalized boson operator $\hat{\alpha}$ and cavity mode frequency $\tilde{\omega}_\mathrm{cav}$ \cite{ciuti2021cavity}. By introducing fermionic operators for the electrons, projecting on the relevant Landau band (the last occupied) and using the basis of electronic single-particle states labeled by the angular momentum $m$, we get  
$
\Hat{\mathcal{H}}
=
\Hat{\mathcal{H}}_{\mathrm{cav}}
+
\Hat{\mathcal{H}}_{\mathrm{coulomb}}
+
\Hat{\mathcal{H}}_{\mathrm{dia}}
$, where 
$\Hat{\mathcal{H}}_{\mathrm{cav}} 
= 
    \hbar \tilde{\omega}_{\mathrm{cav}}
    \hat{\alpha}^{\dagger} \hat{\alpha} $, $
    \Hat{\mathcal{H}}_{\mathrm{coulomb}} 
    = \frac{1}{2} \sum_{mnpq} 
    \langle m,n \vert \hat{v}_{12} \vert q, p \rangle
    \hat{c}^{\dagger}_m \hat{c}^{\dagger}_n \hat{c}_p \hat{c}_q$ and $
    \Hat{\mathcal{H}}_{\mathrm{dia}} 
    = 
\left(
    \hat{\alpha}^{\dagger} + \hat{\alpha}
\right)^2
\Hat{\mathcal{H}}_{\mathrm{dia}}^{\mathrm{el}}$
with 
\begin{align}
    \Hat{\mathcal{H}}_{\mathrm{dia}}^{\mathrm{el}} \, 
    &= \nonumber
    \mathcal{D}_1
    \sum_m 
    \left(
        \sqrt{m+1} \hat{c}^{\dagger}_{m+1} \hat{c}_{m} + \mathrm{h.c.}
    \right) \, \\
    \nonumber
    &+ 
    \frac{\mathcal{D}_2}{2}
    \sum_m 
    \left(
        \sqrt{m+1} \sqrt{m+2} \hat{c}^{\dagger}_{m+2} \hat{c}_{m} + \mathrm{h.c.}
    \right) \\
    &+ 
    \mathcal{D}_2
    \sum_m 
    m \; \hat{c}^{\dagger}_{m} \hat{c}_{m}  
    \, ,
\end{align}
being
$
\mathcal{D}_1
=
\left(
    \frac{ \omega_{\mathrm{cav}} }{ \tilde{\omega}_{\mathrm{cav}} }
\right)
\sqrt{2} A_{\mathrm{vac}} \mathcal{G}_{\mathrm{A}}  \ell 
\left(
    \frac{e^2}{2m_{\star}}
\right)
$
and
$
\mathcal{D}_2
=
\left(
    \frac{ \omega_{\mathrm{cav}} }{ \tilde{\omega}_{\mathrm{cav}} }
\right)
\left(
     \mathcal{G}_{\mathrm{A}}
\ell \right)^2
\left(
    \frac{e^2}{2m_{\star}}
\right)
$, where $\ell = \sqrt
{\hbar/eB}$ is the magnetic length.
Note that the gradient is essential to create an intra-band interaction. If the gradient of the field is zero, photons cannot be exchanged between electrons in the considered Landau band.  The fermionic operator $c^{\dagger}_m$ creates an electron in the state with angular momentum $m$ in the lowest Landau level. For simplicity, we are omitting the spin index. 

Considering processes with up to $2$ exchanged virtual cavity photons, we can adiabatically eliminate the cavity photon field \cite{Cohen_Tannoudji_atomphoton} and obtain an effective electronic Hamiltonian \cite{borici_in_preparation} :
\begin{equation}
    \Hat{\mathcal{H}}_{\mathrm{eff}}
    =
    \Hat{\mathcal{H}}_{\mathrm{dia}}^{\mathrm{el}}
    -
    \frac{1}{\hbar \tilde{\omega}_{\mathrm{cav}}}
    \left(
        \Hat{\mathcal{H}}_{\mathrm{dia}}^{\mathrm{el}}
    \right)^2 \, .
    \end{equation}
The term 
$
\hat{{\mathcal H}}^{(\mathrm{cav})}_{e-e} = - 
    \frac{1}{\hbar \tilde{\omega}_{\mathrm{cav}}}
    \left(
        \Hat{\mathcal{H}}_{\mathrm{dia}}^{\mathrm{el}}
    \right)^2
  $  contains products of four fermionic operators and is, therefore, the cavity-mediated electron-electron interaction, which is attractive due to the minus sign.    

As shown in the following, remarkably, we have been able to calculate analytically the effect of $\hat{{\mathcal H}}^{(\mathrm{cav})}_{e-e}$  on the exchange splitting at odd integer filling factors (when the 2D electron gas is spin polarized) and also the modification of fractional quantum Hall gaps in the so-called Single Mode Approximation \cite{Girvin_master_piece}. 

 The cavity contribution to the exchange splitting
is computed by considering the matrix elements of 
$
\langle \mathrm{FP} \vert
\hat{{\mathcal H}}^{(\mathrm{cav})}_{e-e}
\vert \mathrm{FP} \rangle
$, 
on the fully packed state 
$
\vert \mathrm{FP} \rangle
=
\hat{c}^{\dagger}_0 ... \hat{c}^{\dagger}_{N_{\mathrm{deg}} - 1} 
\vert \mathrm{vac} \rangle
$, which corresponds to an odd integer filling factor. Such matrix elements can be calculated exactly. When the Landau degeneracy $N_{\mathrm{deg}} \gg 1$,  
the term depending on $\mathcal{D}_2^2$ is the dominant one so that the many-electron exchange energy  in the Landau band reads $
    \nonumber
    \Delta E^{\mathrm{exc}}
    =
    \langle \mathrm{FP} \vert
\hat{\mathcal{H}}^{(\mathrm{cav})}_{e-e}
\vert \mathrm{FP} \rangle
    \simeq
    \frac{\mathcal{D}_2^2}{12 \hbar \tilde{\omega}_{\mathrm{cav}}}
    N_{\mathrm{deg}}^3 \, 
$, where $N_{\mathrm{deg}} = \frac{L_x L_y}{2 \pi \ell^2}$ with $L_x$ and $L_y$ being the spatial dimensions of the 2D sample.
Since the effective g-factor is dominated by Coulomb interaction and the cavity-mediated potential is attractive, the cavity correction to $g^{\star}$ (the absolute value of g-factor) is hence negative:
\begin{equation}
    \Delta g^{\star}
    = -
    \frac{1}{N_{\mathrm{deg}}} 
    \frac{\Delta E^{\mathrm{exc}}}{\mu_B B}
     \simeq - 
\frac{1}{48}
    \frac{e^4}{\hbar \mu_B B m_{\star}^2}
    \frac{1}{\tilde{\omega}_{\mathrm{cav}}^5}  
        \mathcal{G}_{\mathrm{E}}^4 \ell^4 N_{\mathrm{deg}}^2
    \, ,\end{equation}
where $\mu_B$ is the electron Bohr magneton. Let us consider parameters corresponding to our experimental configuration, namely $\tilde{\omega}_{\mathrm{cav}} = 7.5 \cross 10^{11} \mathrm{Hz/rad}$, a density of electrons $n = 2 \cdot 10^{11} {\mathrm{cm}}^{-2}$,  
$L_x = L_y = 100 \cross 10^{-6} \mathrm{m}$,
$B= 0.5 \mathrm{T}$, and with the spatial gradient of the vacuum electric field being 
$\mathcal{G}_{\mathrm{E}} =   4 \times 10^8 \mathrm{V/m}^2$. With these values, we get $\Delta g^{\star}_2 \simeq -2$ that is extremely close to what is observed in the experiment.  Note that this is a simplified theory that considers a single cavity mode with a constant spatial gradient. Higher-order spatial derivatives of cavity vacuum field will provide corrections that scale with higher powers of the number of electrons. However, it already provides an impressive agreement with only one fit parameter, the vacuum field spatial gradient $\mathcal{G}_{\mathrm{E}}$, whose order of magnitude is well consistent with our numerical calculations of the cavity mode spatial profile.

In order to describe fractional quantum Hall states, for any two-body interaction 
% $\hat{V}$ 
that conserves angular momentum, the Haldane pseudo-potential components are key quantities. 
They are defined as the expectation values of the interaction Hamiltonian on two-body states with relative angular momentum $m$ \cite{Girvin_2019}.
For the bare and repulsive Coulomb electron-electron interaction, the Haldane pseudopotentias read
$
 v_{m}^{(\mathrm{C})}
\simeq
\left(
    \frac{e^2}{4 \pi \epsilon_0 \epsilon_r \ell}
\right)
\frac{1}{2\sqrt{m}} \, 
$, where $\epsilon_0$ is the vacuum permittivity and $\epsilon_r$ the semiconductor dielectric constant. Our cavity-mediated electron-electron interaction
$\hat{\mathcal{H}}^{(\mathrm{cav})}_{e-e}$ describes processes of the type $\hat{c}^{\dagger}_{m_1} \hat{c}^{\dagger}_{m_2} \hat{c}_{m_1'} \hat{c}_{m_2'}$ where either the angular momentum is conserved ($m_1 + m_2 = m_1' + m_2'$) or changed by two ($m_1 + m_2 = m_1' + m_2' \pm 2$). Note that there is also the possibility of a unity change  ($m_1 + m_2 = m_1' + m_2' \pm 1$), but this corresponds to the $\mathcal{D}_1$-term that can be neglected with respect to the $\mathcal{D}_2$-term for a macroscopic number of electrons. In such large particle number limit, where the involved angular momenta are large, we can approximate the interaction as conserving angular momentum. We have found the following long-range Haldane pseudo-potential for the cavity-mediated attractive interaction
$
    v_{m}^{\mathrm{(cav)}}
    =
   (
        -\frac{\mathcal{D}_2^2}{8 \hbar \tilde{\omega}_{\mathrm{cav}}}
    )
    \left(
        m^2 - m
    \right) \, .
    \label{cavityHaldane}
$ An interaction potential that contains the same Haldane components as the cavity-mediated one that we derived is:
$
    V^{(\mathrm{cav})} ( r )
    =
    \left(
        -
        \frac{
            \mathcal{D}_2^2     
        }{
            8 \hbar \tilde{\omega}_{\mathrm{cav}}
        }
    \right)
    \left[
        \frac{1}{16}
        \left(
            \frac{r}{\ell}
        \right)^4
        - 
        \left(
            \frac{r}{\ell}
        \right)^2
        +
        2
    \right] 
$. 
Note that the divergence at large distances $r$ is due to the approximation of considering a cavity mode with a constant spatial gradient $\mathcal{G}_{\mathrm{E}}$.  A cutoff at large distances is not only expected by the finite-size nature of the Hall bar, but also due to the fact that the spatial region with large gradients is a fraction of the sample (see Fig. \ref{fig1}).
In order to evaluate the many-body gap of fractional quantum Hall states, a powerful technique is the magneto-roton theory \cite{Girvin_master_piece}, which assumes that the collective excitation gap can be determined by the sole knowledge of the (Laughlin) ground state. 
The fractional quantum Hall gap in the presence of the cavity and Coulomb interaction is given by the formula:
\begin{equation}
\Delta^{(\mathrm{cav}+ \mathrm{C})} = \min_k \frac{\overline{f}^{(\mathrm{cav})}( k ) + \overline{f}^{(\mathrm{C)}}( k ) }{\bar{s}(k)} \, ,
\end{equation}
where $\bar{s}(k)$ is the Fourier transform of the density-density correlation function for the considered ground state, projected onto the lowest Landau level. The numerator is the so-called projected oscillator strength\cite{Girvin_master_piece}, which is a linear functional of the electron-electron interaction   $V = V^{(\mathrm{cav})}  + V^{(\mathrm{C})}$ (cavity and Coulomb contributions, see also our Supplementary Material). Plugging the Fourier transform of our $V^{(\mathrm{cav})} ( r )$ provides a divergent result. We have regularized that by introducing an infra-red cutoff due to the finite size of the sample. Doing so (see Supplementary Material), we get $
\overline{f}^{(\mathrm{cav})}( k ) \simeq
0.04
    \left(
        \frac{L}{\ell}
    \right)^4
    \left(
        \frac{
            \mathcal{D}_2^2
        }{
            8 \hbar \tilde{\omega}_{\mathrm{cav}}    
        }
    \right)
    (\ell k)^2 e^{-\frac{1}{2}(\ell k)^2} 
    \left(
        \frac{
            \nu
        }{
            8 \pi^2
        } \right )\, ,$
where $L$ is the length associated to the infra-red cutoff .  As we have verified explicitly,  this kind of $k$-dependence does not shift significantly the wavevector $k_{\mathrm{min}}$ of the magnetoroton minimum.  Hence, the variation of the gap due to the presence of the cavity can be roughly approximated as:
\begin{equation}
\Delta^{(\mathrm{cav + \mathrm{C} })}  - \Delta^{(\mathrm{C })}  \simeq
    0.04
    \left(
        \frac{L}{\ell}
    \right)^4
        \frac{
            \mathcal{D}_2^2
        }{
            8 \hbar \tilde{\omega}_{\mathrm{cav}}    
        }
    \frac{(\ell k_\mathrm{min})^2 e^{-\frac{1}{2}(\ell k_\mathrm{min})^2} }{\overline{s}({k_\mathrm{min}})}
        \frac{
            \nu
        }{
            8 \pi^2
        }
         \,.
\end{equation}

Note that, as in the case of the cavity-mediated g-factor, the variation of the fractional quantum Hall gap scales with $\mathcal{D}_2^2 L^4\propto \mathcal{G}_{\mathrm{E}}^4  N_{\mathrm{deg}}^2$, showing again the key role of the spatial gradient of the vacuum electric field and the collective contribution of the two-dimensional electron gas due to the long-range nature of the perturbation. 
By taking $L = 100 \si{\micro\meter}$ (roughly the system size) and spatial gradients of the same order as the ones used for the calculation of the exchange splitting, setting $\ell k_{\mathrm{min}} = 1.3$ and $s(k_{\mathrm{min}})\sim1$ (the values for the $1/3$-Laughlin state \cite{Girvin_master_piece}), we get $
\frac{
    \Delta^{(\mathrm{cav + C})}  
    - 
    \Delta^{(\mathrm{C})}
}{
    \Delta^{(\mathrm{C})}
}
\approx 0.5
$, which is comparable to what we observed in our experiments. Although the already complex theory is based on many simplifications (single-mode cavity, constant spatial gradient of the vacuum electric field, magneto-roton theory for the excitation gap), the magnitude of the predictions is consistent with the experimental observations.  

This joint experimental and theoretical study has demonstrated the capacity of electromagnetic vacuum fields with pronounced spatial gradients to control complex, strongly correlated electronic systems, specifically fractional quantum Hall phases. Notably, we observed a significant enhancement of fractional quantum Hall gaps within a key group of fractions. We have revealed that the exchange of virtual cavity photons can induce an effective attractive long-range electron-electron interaction, which competes with the Coulomb interaction. Additionally, we found that this cavity-mediated attractive interaction substantially reduces the exchange spin splitting, which is responsible for the activation energy gap at odd integer filling factors.

These experimental results were achieved using an innovative ``hovering'' resonator technique. This method finely tunes the electromagnetic vacuum fields by adjusting the distance between the cavity and the Hall bar, offering precise control. Moreover, this approach has broader implications and can be generalized to any 2D material. Specifically, leveraging cavity vacuum fields in this manner could manipulate strongly correlated phases in moiré materials, such as by stabilizing fractional Chern insulators\cite{neupert,sheng,regnault} or altering superconductivity \cite{Cao2018} properties.

Note added: while completing this manuscript, we became aware of a theoretical preprint focusing on the resonant hybridization of the magnetoroton with the cavity photon field \cite{Bacciconi2024}.

\bibliography{scibib}% common bib file
\bibliographystyle{apsrev4-1}

\end{document}

\title[Italic]{\textbf{Supplementary Material for:}\\ 
Enhanced fractional quantum Hall gaps in a two-dimensional electron gas  \\ 
coupled to a hovering split-ring resonator}
\author{Josefine Enkner}
\affiliation{Institute for Quantum Electronics, ETH Zurich, CH-8093 Zurich, Switzerland}
\affiliation{Quantum Center, ETH Zürich, 8093 Zürich, Switzerland}

\author{Lorenzo Graziotto}
\affiliation{Institute for Quantum Electronics, ETH Zurich, CH-8093 Zurich, Switzerland}
\affiliation{Quantum Center, ETH Zürich, 8093 Zürich, Switzerland}

\author{Dalin Boriçi}
\affiliation{Université Paris Cité, CNRS, Matériaux et Phénomènes Quantiques, 75013, Paris, France}

\author{Felice Appugliese}
\affiliation{Institute for Quantum Electronics, ETH Zurich, CH-8093 Zurich, Switzerland}
\affiliation{Quantum Center, ETH Zürich, 8093 Zürich, Switzerland}

\author{Christian Reichl}
\affiliation{Laboratory for Solid State Physics, ETH Zürich, 8093 Zürich, Switzerland}

\author{Giacomo Scalari}
\affiliation{Institute for Quantum Electronics, ETH Zurich, CH-8093 Zurich, Switzerland}
\affiliation{Quantum Center, ETH Zürich, 8093 Zürich, Switzerland}

\author{Nicolas Regnault}
\affiliation{Laboratoire de Physique de l’Ecole normale sup\'erieure,ENS, Universit\'e PSL, CNRS, Sorbonne Universit\'e}
\affiliation{Department of Physics, Princeton University, Princeton, NJ 08544, USA}

\author{Werner Wegscheider}
\affiliation{Laboratory for Solid State Physics, ETH Zürich, 8093 Zürich, Switzerland} 

\author{Cristiano Ciuti}
\affiliation{Université Paris Cité, CNRS, Matériaux et Phénomènes Quantiques, 75013, Paris, France}

\author{Jérôme Faist}
\affiliation{Institute for Quantum Electronics, ETH Zurich, CH-8093 Zurich, Switzerland}
\affiliation{Quantum Center, ETH Zürich, 8093 Zürich, Switzerland}


\keywords{keyword1, Keyword2, Keyword3, Keyword4}



\maketitle

This Supplementary Material concerns both the experiments and theory described in the main manuscript. In particular, it provides additional details concerning the setup, sample, and measurement techniques. On the theoretical side, it contains sections showing the negligible role of electrostatic effects, some exact diagonalization results for a small number of electrons, and details about the magneto-roton theory of the fractional quantum Hall gaps that can be applied for extensive systems.

\section{Material and Methods}\label{sec1}

\begin{figure*}[h!]
\centering
\includegraphics[width=0.7\textwidth]{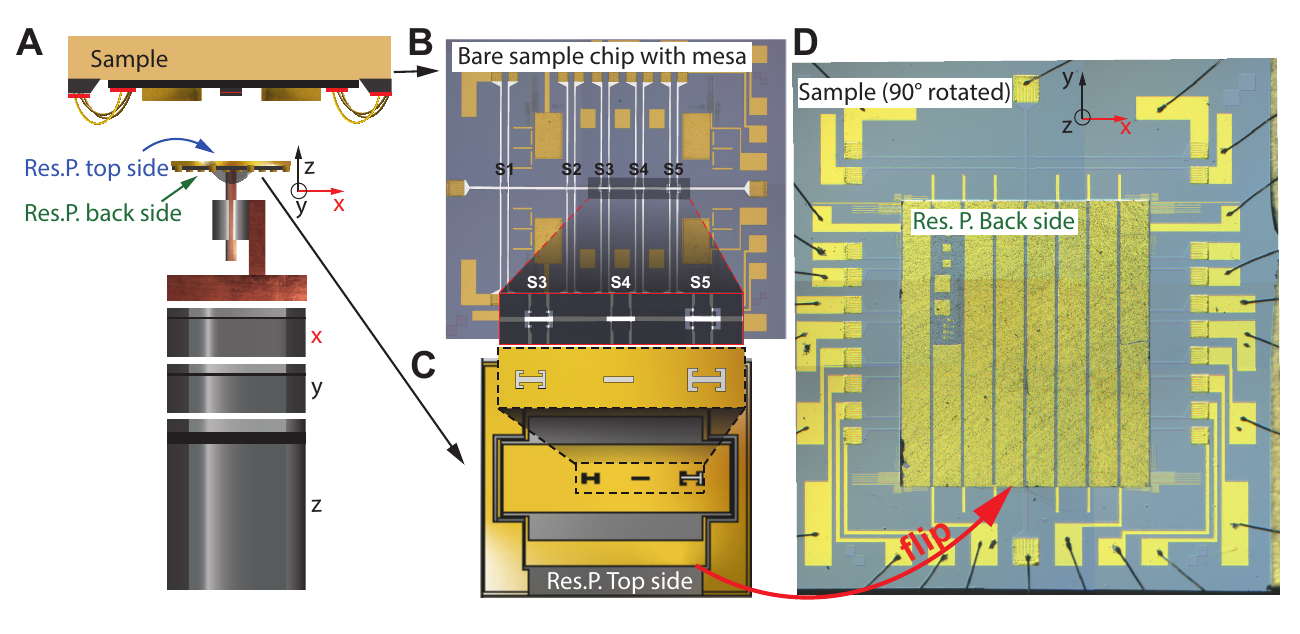}
\caption{Extended setup description: \textbf{(A)} Sketch of the sample chip mounted top down in the cryostat with the movable resonator plane (Res.P.), top side facing up, mounted from below with a copper rod attaching it to three nanopositioners moving in x, y and z direction. \textbf{(B)} Optical microscope picture of the chip with the etched mesa structure (highlighted in white) and all of the Hall bar samples S1 to S5. Zoom in red frame: depiction of how the CSRR on the resonator plane and the Hall bar are designed to overlay. \textbf{(C)} Drawing of the resonator plane design with zoom-in to the CSRR. \textbf{(D)} Optical microscope picture of the pre-aligned sample (90$^{\circ}$ rotated to \textbf{(BC)}): in this view the resonator plane has already been flipped. The thick gold lines on the back side of the resonator plane serve as alignment markers that line up with gold markers on the sample chip.} 
\label{fig:extended_setup}
\end{figure*}

\subsection{Material details on the Hall bar hetero-structure and the resonator plane}

The GaAs-based heterostructure D151202B employed for the Hall bar was grown via molecular beam epitaxy (MBE) by Dr.\ Christian Reichl in the Solid State Laboratory at ETH Zürich. The D151202B hosts a double-side doped, \SI{30}{nm} wide GaAs/AlGaAs quantum well with a \SI{100}{nm} spacer and is buried $233~\si{\nano\meter}$ under the surface. Due to the double-side doped design of the quantum well, the sample is immune to gate-induced density and mobility fluctuations. The material exhibits a high electron mobility $1.69\times 10^7~\mathrm{cm^2 V^{-1}s^{-1}}$ and sheet density $2.06\times 10^{11}~\mathrm{cm^{-2}}$, measured at \SI{1.3}{K} without illumination. The resonator plane consists of a square $3620\times3200~\si{\micro\meter^2}$ piece of insulating GaAs, lapped down to a thickness of $150~\si{\micro\meter}$, on top of which the complementary split-ring resonator (CSRR) is fabricated.

\subsection{Design of the Hall bar sample and the resonator plane}

On a $6.3\times6 \SI{}{mm^2}$ chip of the D151202B material we define by wet etching a $5376~\si{\micro\meter}$-long and \SI{350}{nm}-thick mesa which we probe at 5 different positions S1, S2, S3, S4 and S5 (see Fig.~\ref{fig:extended_setup}B). Each of these Hall bar samples is a minimum of $500~\si{\micro\meter}$ apart from each other. Probes leading to the samples have a length of 2 mm and a width of $30~\si{\micro\meter}$ and connect to the Hall bar with a tapered width of $10~\si{\micro\meter}$. This mask layout allows us to probe multiple Hall bar samples on the same mesa at the same time. Since all samples come from the same chip and share a source and drain, we can rule out density and mobility drifts arising from temperature or another spurious factor. The sample shown in Fig.~\ref{fig:extended_setup}B has four pillar-shaped landing pads (the big gold rectangles between S1 and S2, and after S5 in Fig.~\ref{fig:extended_setup}B) that probe the electrical connection between the resonator plane and the Hall bar sample and serve as safeguard against damaging the Hall bar with the movable hovering resonator plane. Indeed, when the gold resonator plane is in touch with the pillars, it electrically shortens the connection between them, thus allowing the touching position to be confirmed from outside of the cryostat.

On the top side of the resonator plane (Fig.~\ref{fig:extended_setup}C), we lithographically define the three complementary split ring resonators and two electrically disconnected gold planes at the corners, which are electrically connected via evaporated gold on the substrate edges to the back side of the resonator plane chip. As seen in Fig.~\ref{fig:extended_setup}D, where we show a picture of the top view of the pre-aligned Hall bar sample with the resonator plane (here back side view), alignment markers on both the backside of the resonator plane (lifted-off lines) and the Hall bar sample guarantee the spatial overlap of the samples S3, S4, and S5 with the resonators (see bottom Fig.~\ref{fig:extended_setup}B). The uncoupled fundamental mode of CSRR on S3 is engineered to resonate at a frequency of 150$~\si{\giga\hertz}$, for S4 at 200$~\si{\giga\hertz}$, and for S5 at 130$~\si{\giga\hertz}$. The resonator gaps of the CSRRs overlapping with S3 and S4 have a width of $40~\si{\micro\meter}$, making alignment precision a crucial factor, while the gap of the CSRR overlapping with S5 spans a width of 50$~\si{\micro\meter}$. 

\begin{figure*}[h!]
\centering
\includegraphics[width=0.8\textwidth]{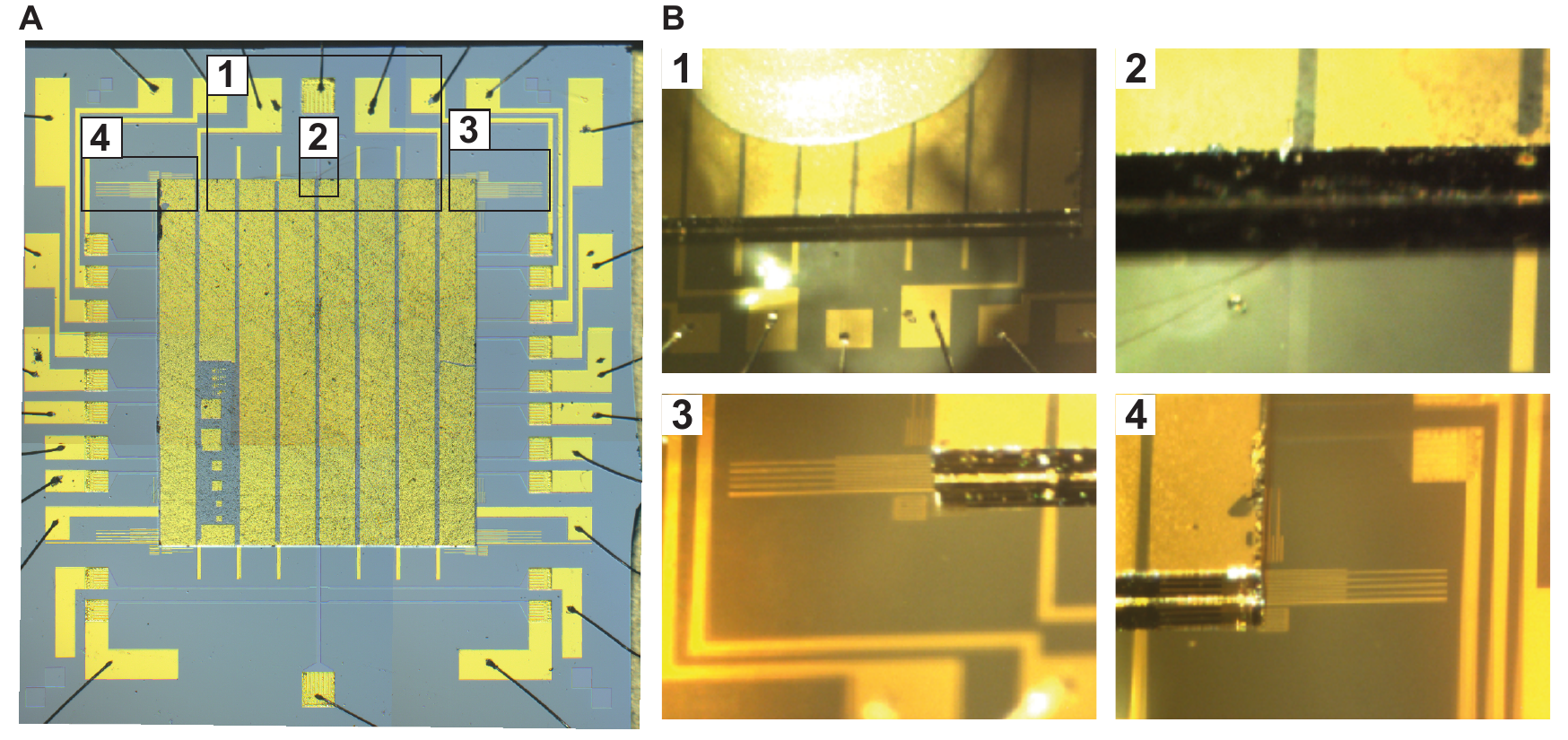}
\caption{Close-up photographs detailing the alignment: \textbf{(A)} Optical microscope picture of pre-aligned structure: sample chip below the upside-down resonator plane. \textbf{(B)} Digital camera pictures of the alignment after mounting on the cryostat cold finger. In B1, one can see the glue with which the copper rod gets attached to the back side of the resonator plane. B2 shows the precise alignment of the 40 $\si{\micro\meter}$ wide Hall bar to the marker to the back of the resonator plane. B3 and B4 show close-up pictures on the edges of the tip where with alignment markers forming a grid, last precise alignment adjustments can be made. The angular alignment is also ensured via this procedure.} 
\label{fig:closeups}
\end{figure*}

\subsection{Fabrication of the Hall bar sample and the resonator plane}
\textbf{Hall bar sample:} Starting from the heterostructure, grown by MBE, the fabrication of the sample is carried out by optical lithography. The steps are the following:

\begin{itemize}
    \item Definition of the Hall bar using positive resist (AZ1505) and photolithography;
    \item Etching of the mesa using a very diluted Piranha acid (\ce{H2SO4}:\ce{H2O2}($30\%$):\ce{H2O} 1:8:60);
    \item Contacts are defined photolithographically using a negative image-reversal resist (AZ5214E);
    \item Ge/Au/Ge/Au/Ni/Au (41/84/41/84/63/40 nm) eutectic mixture evaporation, lift off and annealing (500$^{\circ}$C for \SI{300}{s});
    \item Extra gold pads are defined again by optical lithography with image-reversal resist;
    \item Deposition of Ti/Au (11/200 nm) and lift off.
\end{itemize}
\textbf{Resonator Plane:} Starting from a $1\times1$ \SI{}{cm^2} piece of insulating GaAs, we define 9 resonator plane structures that, during the process, will get cleaved and processed separately:

\begin{itemize}
    \item Definition of the cleaving guides using a positive resist (AZ1505) and photo-lithography;
    \item Etching of the cleaving guides using a very diluted Piranha acid (\ce{H2SO4}:\ce{H2O2}($30\%$):\ce{H2O} 1:8:60);
    \item Resonator plane is defined photolithographically using a negative image-reversal resist (AZ5214E);
    \item Deposition of Ti/Au (11/200 nm) and lift off;
    \item Spinning protective layer of resist;
    \item Lapping the sample down to a thickness of 150$~\si{\micro\meter}$;
    \item Definition of the alignment markers using a negative resist (AZ5214E) and photo-lithography on the back of the chip; 
    \item Deposition of Ti/Au (11/200 nm) and lift off;
    \item Spinning protective layer of resist on both sides;
    \item Cleaving the GaAs sample into 9 separate resonator planes along the cleaving guides;
    \item Deposition of Ti/Au (11/200 nm) on the side walls and lift off.
\end{itemize}

\begin{figure*}[t!]
\centering
\includegraphics[width=0.7\textwidth]{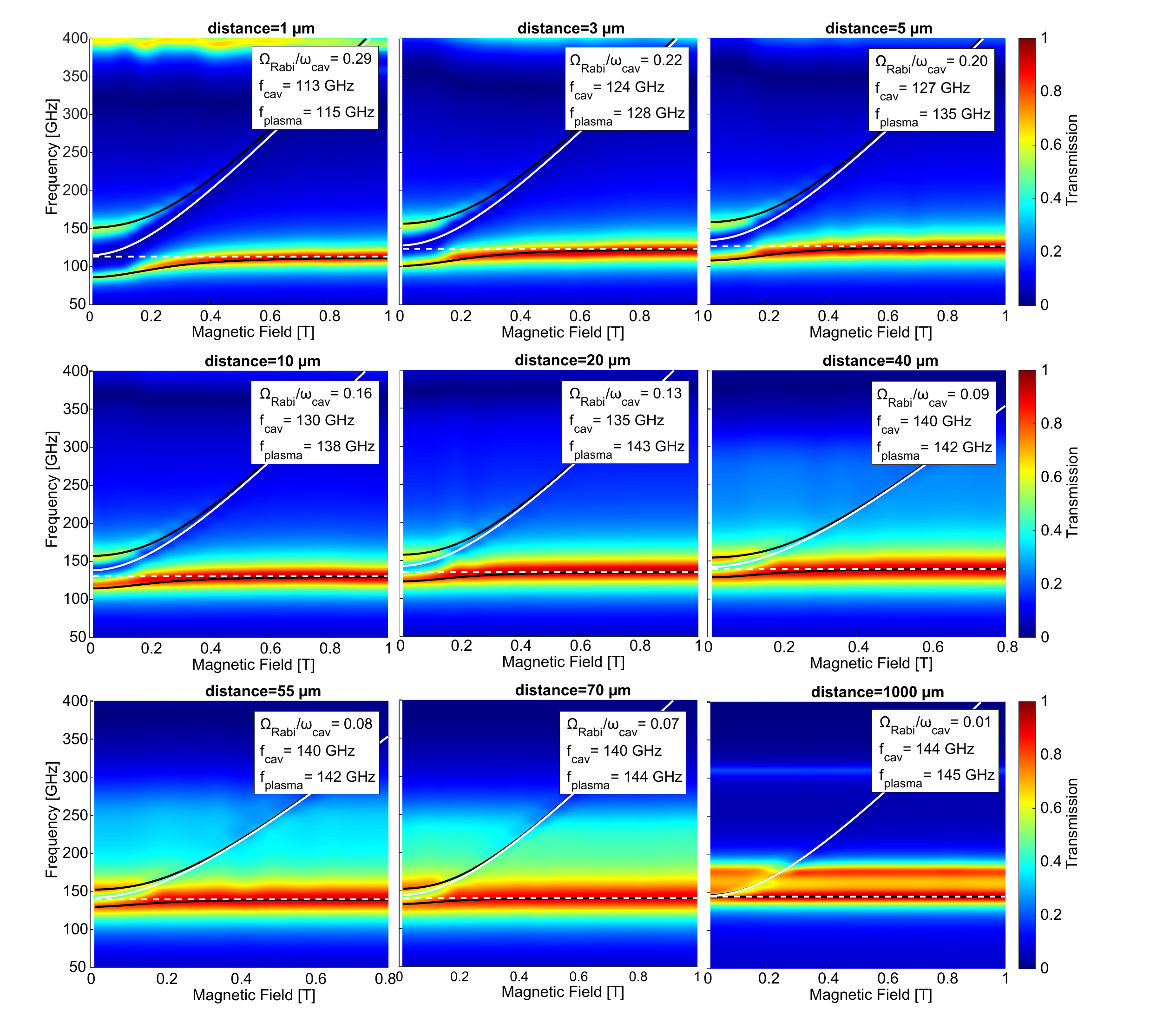}
\caption{Simulation of the polaritonic energy dispersions for different distances between resonator plane and the 2DEG.} 
\label{fig:polariton_simulation}
\end{figure*}

\subsection{Alignment of the Hall bar with the cavity}
A critical factor of this experiment is the correct alignment of the resonator plane with the Hall bar samples defined on the mesa structure on the sample chip. Before transferring the chip into our Bluefors dilution refrigerator, we pre-align the sample and the resonator plane: First, we put a droplet of water on the surface of the Hall bar sample, then put the resonator plane (with the CSRRs facing the Hall bar) on top of the droplet, and then align the resonator plane using a micromanipulator. Finally, we gently press the resonator plane on top of the sample chip so that the Van der Waals force holds the resonator plane in place. Then, the pre-aligned geometry, as shown in Fig.~\ref{fig:closeups}A, can be transferred onto the cold finger in the dilution fridge. At room temperature, the resonator plane is glued to a copper rod, which is mounted to a stack of piezoelectric attocube nanopositioners, which can move in the x, y, and z directions. After the glue is dry, we perform the last alignment procedures before cooling it down. In Fig.~\ref{fig:closeups}B, we show close-up pictures of the alignment of the markers of the resonator plane and sample chip, already mounted on the dilution fridge cold finger at room temperature. Any residual water will be removed by pumping during the cool-down process.

\subsection{Conducting the experiment}

This study utilizes state-of-the-art techniques for analyzing the quantum Hall effect in two-dimensional electron systems~\cite{baer2015transport}. Using a Bluefors dilution refrigerator, we can cool to temperatures as low as \SI{19}{mK}. Voltage measurements are carried out using Zurich Instruments MFLI digital lock-in amplifiers. To inject current symmetrically into the mesa structure, an AC-modulated voltage of \SI{2}{V} (rms) at a demodulation frequency of 2.333 Hz is applied across two 100$~\si{\mega\ohm}$ resistors in series with the mesa, generating a current of \SI{10}{nA} (rms). Differential AC low-noise pre-amplifiers are used to amplify the signal by a factor of 1000 before it reaches the lock-in voltage input. A fourth-order low-pass filter with a time constant of \SI{1.0}{s} is employed to demodulate the input signal to the lock-in amplifier. Additionally, low-pass \SI{100}{kHz} filters are installed before the sample contacts to reduce electrical spikes or potential heating effects from the measurement setup.\\
As shown in Fig.~\ref{fig:extended_setup}A, the three piezoelectric actuators can be moved in order to tune the coupling between the resonator plane and each Hall bar sample. In this study, we only conducted thorough measurements on the samples S4 and S5. Right after cooling down, the resonator plane is brought into contact with the pillar-shaped landing pads on the chip. When a short between all four pillars is measured, we can guarantee a parallel alignment of the resonator plane to the surface of the Hall bar sample, an initial distance of $\approx 450$ nm from the 2DEG, and the equilibrium of the same potential of resonator plane and Hall bar sample. In order to convert the number of retraction steps of the piezoelectric actuators from this initial position, we measure the distance between 0 and 200 retracted steps of the piezoelectric actuators at \SI{4}{K} as the position reading of the piezoelectric actuators is limited at mK-temperatures. We point out that indeed thermal expansion might shift the precision of the estimated position.\\
The setup permits us to measure the longitudinal and transverse resistance of samples S4 and S5 at the same time. Each trace of longitudinal and transverse voltage was measured for a specific distance between the surface of the resonator plane and the surface of the Hall bar as a function of magnetic flux density, which is tuned via a superconductive solenoid magnet capable of reaching \SI{12}{T}. We point out that for the measurement performed at a distance $\SI{0.35}{\micro\meter}$, the Hall bar sample and the resonator plane are not touching and are electrically disconnected from each other.

\section{Finite-element simulations}

\begin{figure}[h!]
\centering
\includegraphics[width=0.5\textwidth]{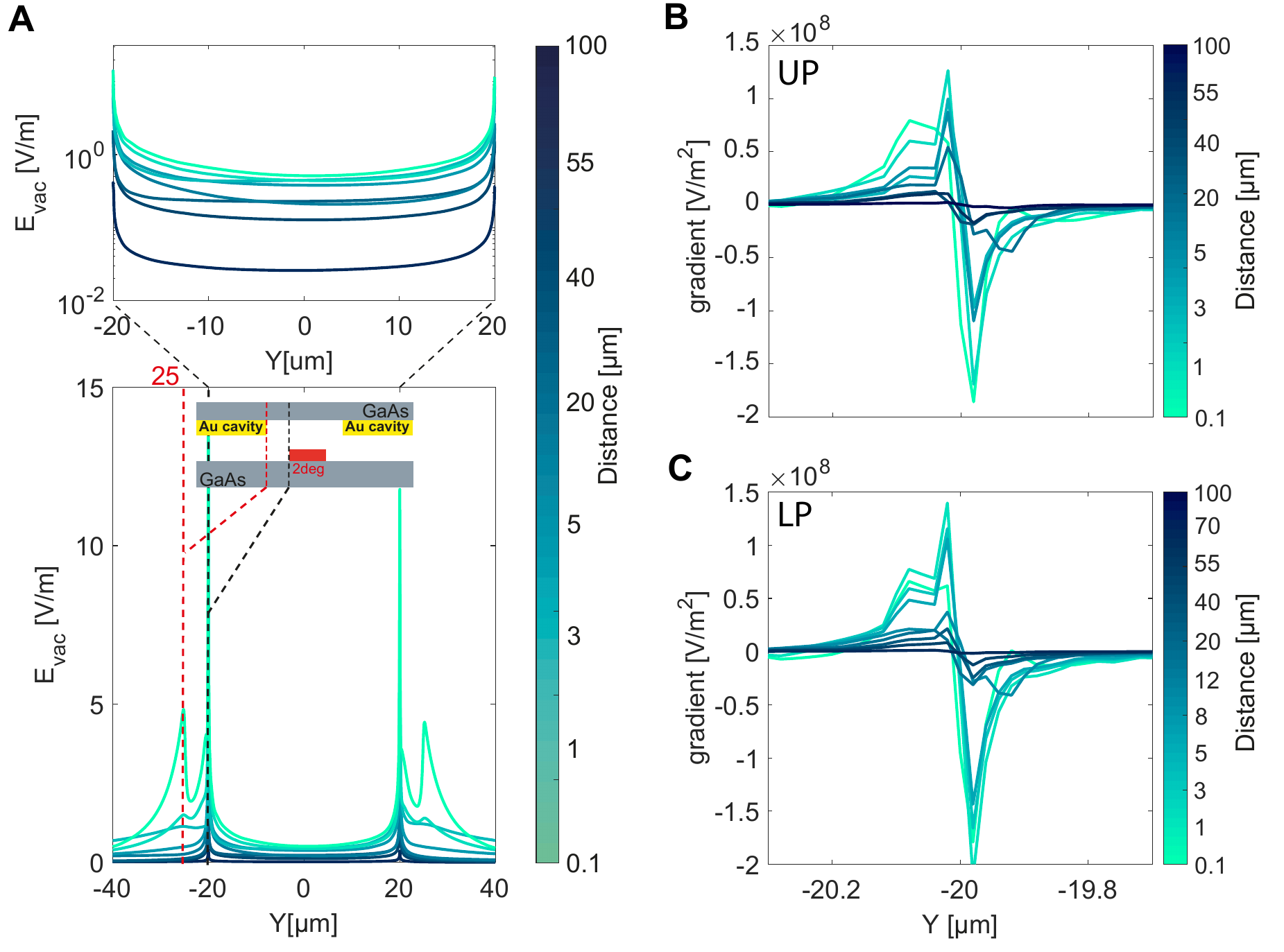}
\caption{Field profile of the upper polariton and field gradients: \textbf{(A)} Field profile of the Vacuum electric field across the 2DEG. Top: Zoom to the area within the Hall bar. \textbf{(B)} Field gradient $\mathcal{G}_A$ of the upper polariton (UP - top) and the lower polariton (LP - bottom)} 
\label{fig:fieldprofile_UP}
\end{figure}

In the following we discuss supporting material on the finite-element simulations performed in order to estimate the normalized coupling $\Omega_\mathrm{Rabi}/\omega_\mathrm{cav}$~\cite{hagenmuller2010ultrastrong} of the system and the field profile of the vacuum electric field $E_\mathrm{vac}$ inside the Hall bar as a function of distance between the 2DEG and the CSRR. All the simulations of Fig.~\ref{fig:polariton_simulation} and Fig.~\ref{fig:fieldprofile_UP} are performed using the CST Microwave Studio software.

The resonator plane is modeled using the standard lossy metal gold from the material library on top of a GaAs substrate. Similarly, the substrate of the Hall bar sample is a block of GaAs. The 2DEG stripe is modeled using a gyrotropic material with bias (i.e., the magnetic field) in the direction perpendicular to the surface. An effective layer thickness was used in order to reduce computational cost. The distance between the 2DEG and the CSRR is tuned parametrically. In Fig.~\ref{fig:polariton_simulation}, we show the polaritonic dispersion obtained from the s-parameter parallel to the excitation field (perpendicular to the resonator gap) from the simulations for different values of the distance between the 2DEG and the resonator plane ($[\si{\micro\meter}]$). Fitting the Hopfield model~\cite{Hopfield} on top of the dispersion, we estimate the normalized coupling that is also used at a later point to calculate the vacuum electric field penetrating the Hall bar. This method tends to overestimate the coupling strength but offers a reliable verification of the expected trend for the resonators, which have been measured optically using our THz time-domain spectroscopy set-up described in Refs.~\cite{paravicini2019magneto,paravicini2017gate,appugliese2022breakdown}.

The normalized coupling strength depends on the effective cavity volume $V_\mathrm{cav}$ and on the number of coupled electrons as $\Omega_\mathrm{Rabi}/\omega_\mathrm{cav}\propto \sqrt{N_\mathrm{2DEG}/V_\mathrm{cav}}$. To estimate the component of the vacuum electric field orthogonal to the Hall bar, we assume the area coupled to the resonator field to be constant as a function of distance and then normalize the electric field $E=\sqrt{E_x^2+E_y^2+E_z^2}$ to the Rabi frequency $\Omega_\mathrm{Rabi}$, estimated through the simulations shown in Fig.~\ref{fig:polariton_simulation}, with
$$E_\mathrm{vac}=E\cdot\sqrt{\left(\frac{\mathcal{E}_\mathrm{vac}^2\cdot W}{\int_{a_1}^{a_2}E^2}\right)} $$
where \(\mathcal{E}_\mathrm{vac}=\frac{\hbar\Omega_\mathrm{Rabi}}{\delta \sqrt{N_e}}\)~\cite{hagenmuller2010ultrastrong}, \(W\) indicates the Hall bar's width, $a_{1,2}$ are the boundaries of the Hall bar, \(\delta\) is the dipole moment, and \(N_e\) corresponds to the number of dipoles.

In Fig.~\ref{fig:fieldprofile_UP} we show the field profile simulation of the upper polariton (UP) within the Hall bar. Simulations were performed on the coupled system at a field corresponding to $B=\SI{0.3}{T}$. Similarly to the field profile shown in the main text, we can again identify four peaks symmetrically centered around zero. Two coincide with the edges of the cavity gap defined by the CSRR and two mark the boundaries of the Hall bar. We point out that the amplitude of the peaks at $\pm25~\si{\micro\meter}$ is lower than for the lower polariton (LP), which can be explained due to the fact that the lower polariton frequency is closer to the cavity frequency. We note that the constant field component inside the Hall bar (Fig. \ref{fig:fieldprofile_UP}A top) is of the order of $E_\mathrm{vac}=0.8$ V when the cavity is close and then decreases with increasing distance. We point out, however, that the gradient $\mathcal{G}_A$ for both lower and upper polaritons is of the order of $10^8$ V/m$^2$, which is in accordance with theoretical predictions. 

\begin{figure}[h!]
\centering
\includegraphics[width=0.5\textwidth]{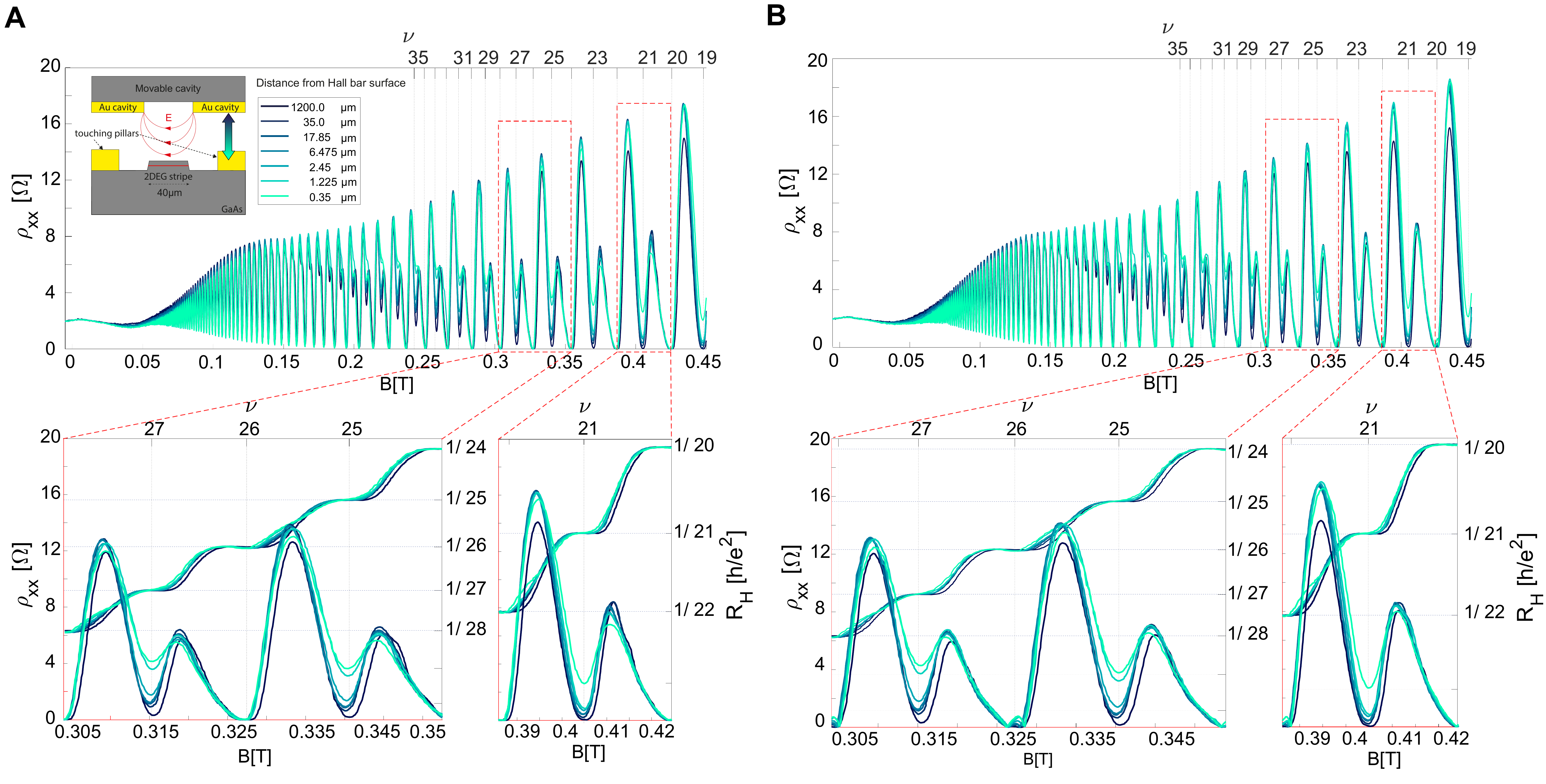}
\caption{S4 data: \textbf{(A)} Data shown for the sample S4 (side 1 - Vxx1) taken at various distances between the cavity and Hall bar. Top: Focus on low field, showing the absence of density and mobility drifts. Bottom: Zoom in to filling factors 27 to 25 and 21. \textbf{(B)} Data shown for sample S4 but on the opposite side (Vxx2) of the Hall bar.} 
\label{fig:S4data}
\end{figure}

\section{Additional Data}

In this section, we discuss the data taken on the second Hall bar S4 coupled to a slot antenna resonator with a resonance frequency of $\omega_\mathrm{slot}=2\pi\times200~\si{\giga\hertz}$ with a resonator gap of 40 $\si{\micro\meter}$. The maximal coupling for this resonator is estimated to be lower than for the CSRR in the main text due to the higher resonance.\\
When measuring the longitudinal transport on both sides of the S4 Hall bar (see Fig.~\ref{fig:S4data}), analogously to the data presented in the main text, we observe the lifting of the minima and a worse quantized plateau in the transverse direction with decreasing distance between the resonator plane and the Hall bar sample. Again, we can already observe a hint of the reduction in the Zeeman split states as the peaks between the odd integer filling factors tend to move closer together.

\begin{figure*}[t!]
\centering
\includegraphics[width=0.9\textwidth]{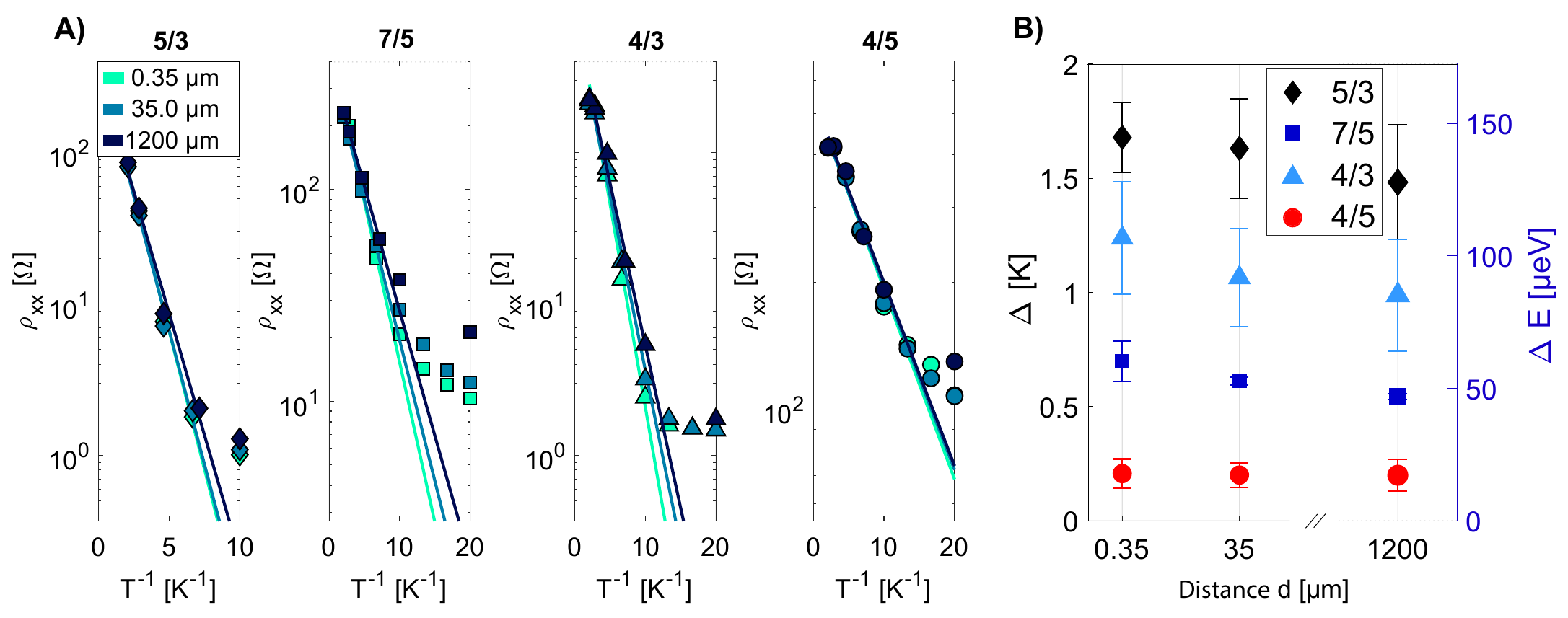}
\caption{Sample S4: \textbf{A)} Arrhenius activation plots for different fractional filling factors for the three distances $0.35~\si{\micro\meter}$, $35~\si{\micro\meter}$ and $1200~\si{\micro\meter}$. \textbf{B)} Extracted activation energy gaps for fractional fillings $5/3$, $7/5$, and $4/3$ as a function of the distance between the split-ring resonator and the Hall bar.} 
\label{fig:fractions_S4}
\end{figure*}

\begin{figure}[h!]
\centering
\includegraphics[width=0.5\textwidth]{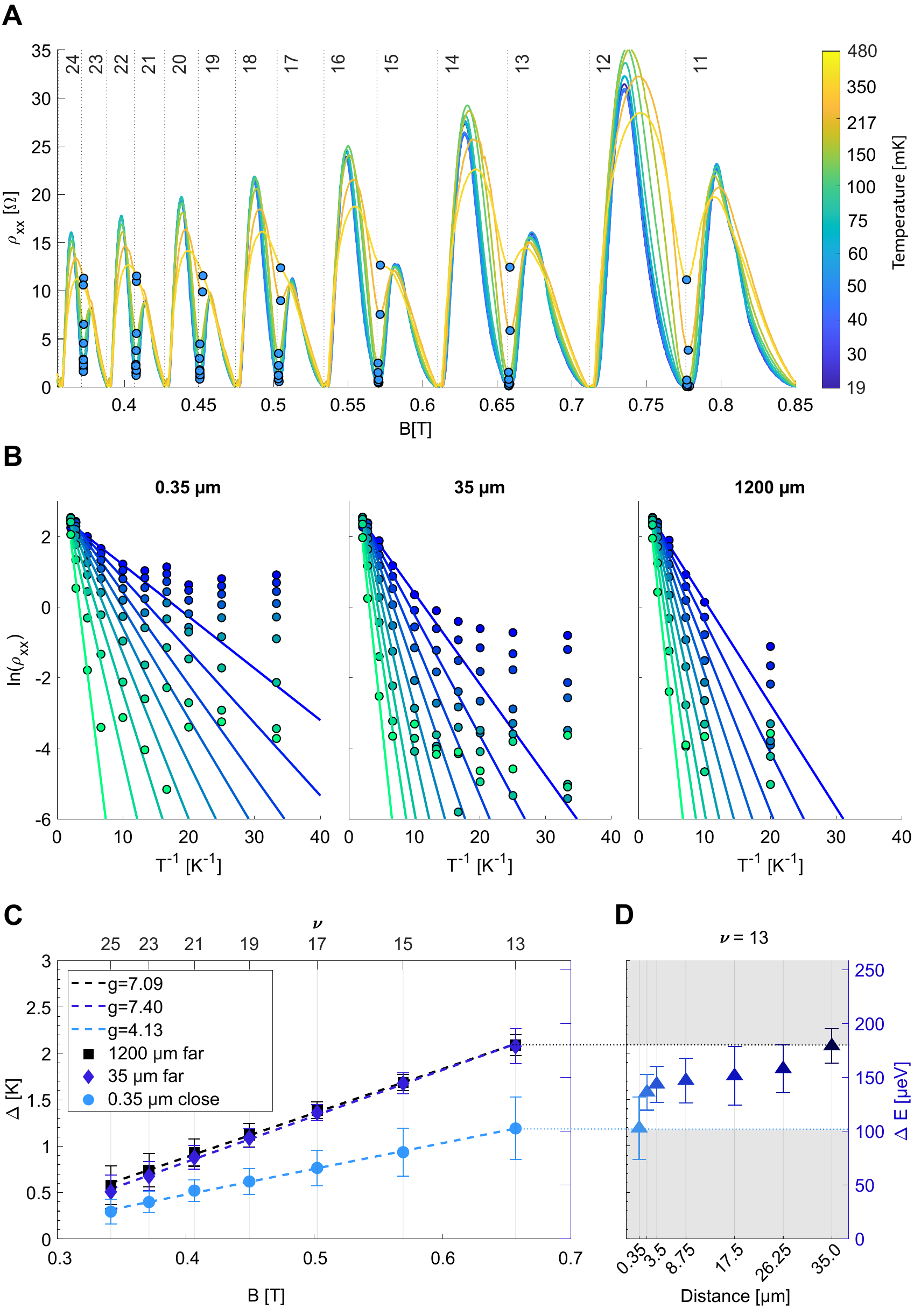}
\caption{Activation and g factor S4: \textbf{(A)} longitudinal data taken at multiple temperatures ranging from 19-480 mK. The blue dots mark the position at which the value for the resistivity was extracted. \textbf{(B)} Arrhensious plots for the temperature sweeps taken at different positions 0.35 $\si{\micro\meter}$, 35 $\si{\micro\meter}$ and 1200 $\si{\micro\meter}$. \textbf{(C)} Extracted activation energies for odd filling factors 25 to 13 as a function of the magnetic field. Dashed lines indicated the fit performed in order to estimate the effective g factor. \textbf{(D)} Evolution of the activation energy of filling factor 13 as a function of distance.} 
\label{fig:S4activation}
\end{figure}

Similarly to the data presented on Fig.~3 in the main text, we observe a reduction of the effective g factor as a function of the distance $d$, as shown in Fig.~\ref{fig:S4activation}C. In Fig.~\ref{fig:S4activation}A, we show the extraction of the points at different temperatures, whose exponential decay as a function of $1/T$ gives the activation energy of the odd integer states (see Fig.~\ref{fig:S4activation}B). In addition, in Fig.~\ref{fig:fractions_S4}A, we show the Arrhenius plots as a function of distance for the fractional states 5/3, 4/3, and 7/5 and 4/5.  In accordance with the findings presented in the main text, we note an enhancement of the fractional states 5/3, 4/3, and 7/5 shown in Fig.~\ref{fig:fractions_S4}B.
\begin{figure}[b!]
\centering
\includegraphics[width=0.5\textwidth]{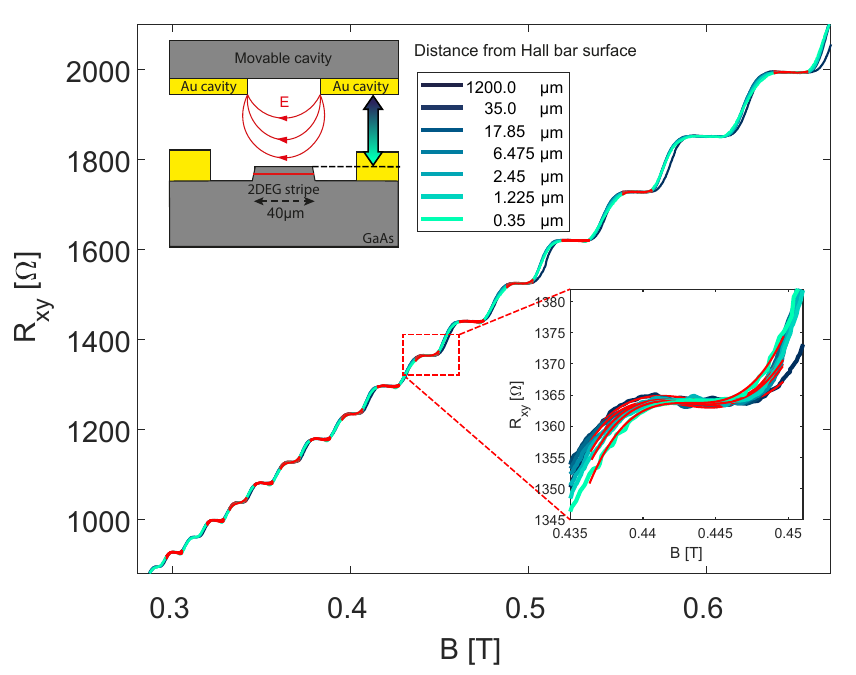}
\caption{Plateau fit S5: Transverse data measured on Hall bar S5 for multiple positions and the polynomial fit (red) around the plateau regions. Inset: Zoom to the plateau of filling factor 22 and the fit in red on top.} 
\label{fig:plateausS5}
\end{figure}

\subsection{Plateau Analysis}
To assess the influence of the cavity on the quality of plateau quantization, we conducted the following analysis, as illustrated in Fig.~\ref{fig:plateausS5}. Each plateau was modeled using a third-order polynomial fit $f(x)=p1\cdot x^3 + p2\cdot x^2 + p3\cdot x + C$. The differences in fits for various distance points are observable in the inset of Fig.~\ref{fig:plateausS5}.

\begin{figure}[h!]
\centering
\includegraphics[width=0.5\textwidth]{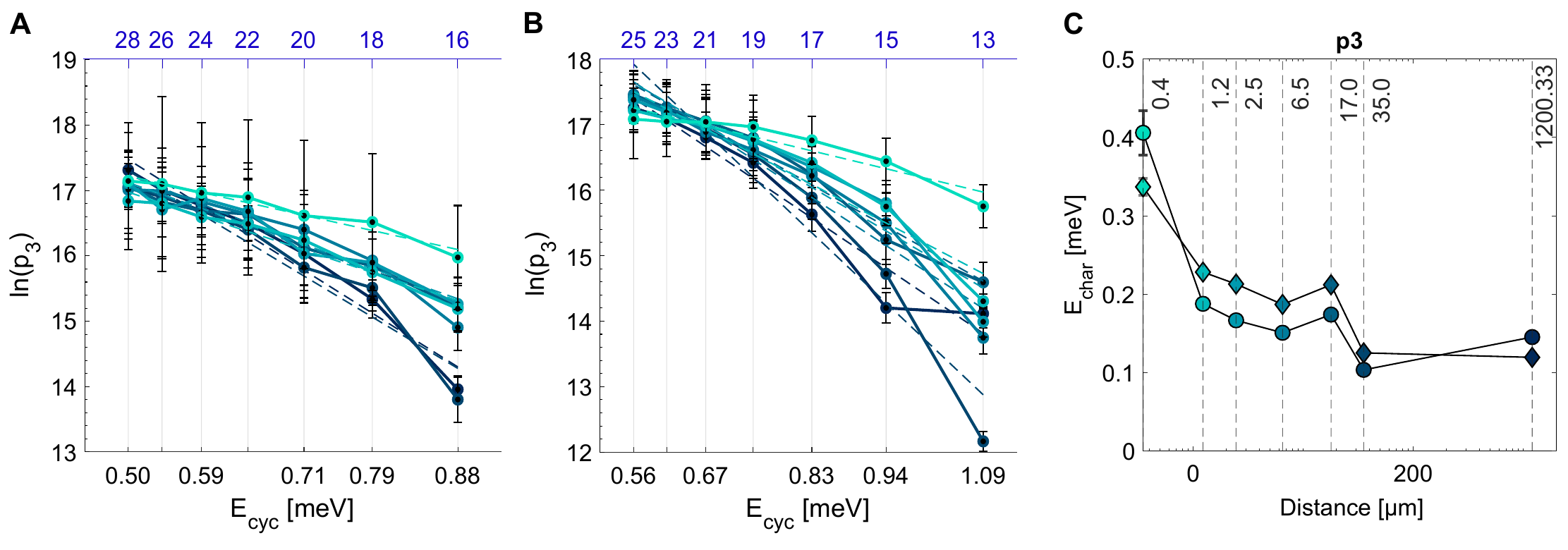}
\caption{Exponential decay of the linear fitting parameter p3: \textbf{(A)} Exponential decay of the linear fitting parameter p3 for even filling factor 28 to 16 as a function of cyclotron energy. Each colored line corresponds to the linear fitting value p3 obtained for a specific distance. Light color is close (0.35 $\si{\micro\meter}$) while dark is far away (1200 $\si{\micro\meter}$). \textbf{(B)} Exponential decay of the linear fitting parameter p3 for odd filling factor 25 to 13 as a function of cyclotron energy. \textbf{(C)} The characteristic energy $E_{char}$, associated with the exponential decay of p3 as a function of $E_{cyc}$, as a function of distance.} 
\label{fig:linear_fitting_parameter}
\end{figure}

Subsequently, we graphed each polynomial component against the cyclotron energy, effectively plotting it as a function of the filling factor. This approach not only revealed a function that appears to decay exponentially but also showed a correlation between the decay rate and the distance at which the measurements were taken (see Fig.~\ref{fig:linear_fitting_parameter}AB). The observed exponential decay in this study mirrors the results found in~\cite{appugliese2022breakdown}, which documented a similar exponential decrease in cavity-induced resistivity. In that work, the characteristic energy of the exponential decay was identified as $E_\mathrm{char}=0.4~\si{\milli\eV}$. Here, we look at the linear component of the polynomial fit p3 as a function of cyclotron energy. Our analysis identifies a characteristic energy ($E_\mathrm{char}$) of $0.4~\si{\milli\eV}$ when the resonator plane is proximate, and $0.1~\si{\milli\eV}$ when it is distant (see Fig.~\ref{fig:linear_fitting_parameter}C). This analysis not only allows us to quantify the impact of the cavity on both odd and even plateaus but also allows us to connect our work to the previous findings of Ref.~\cite{appugliese2022breakdown}.

\section{Negligible role of electrostatic screening on Coulomb potential from distant hovering resonator}

\begin{figure}[h!]
\includegraphics[width=0.5\textwidth]{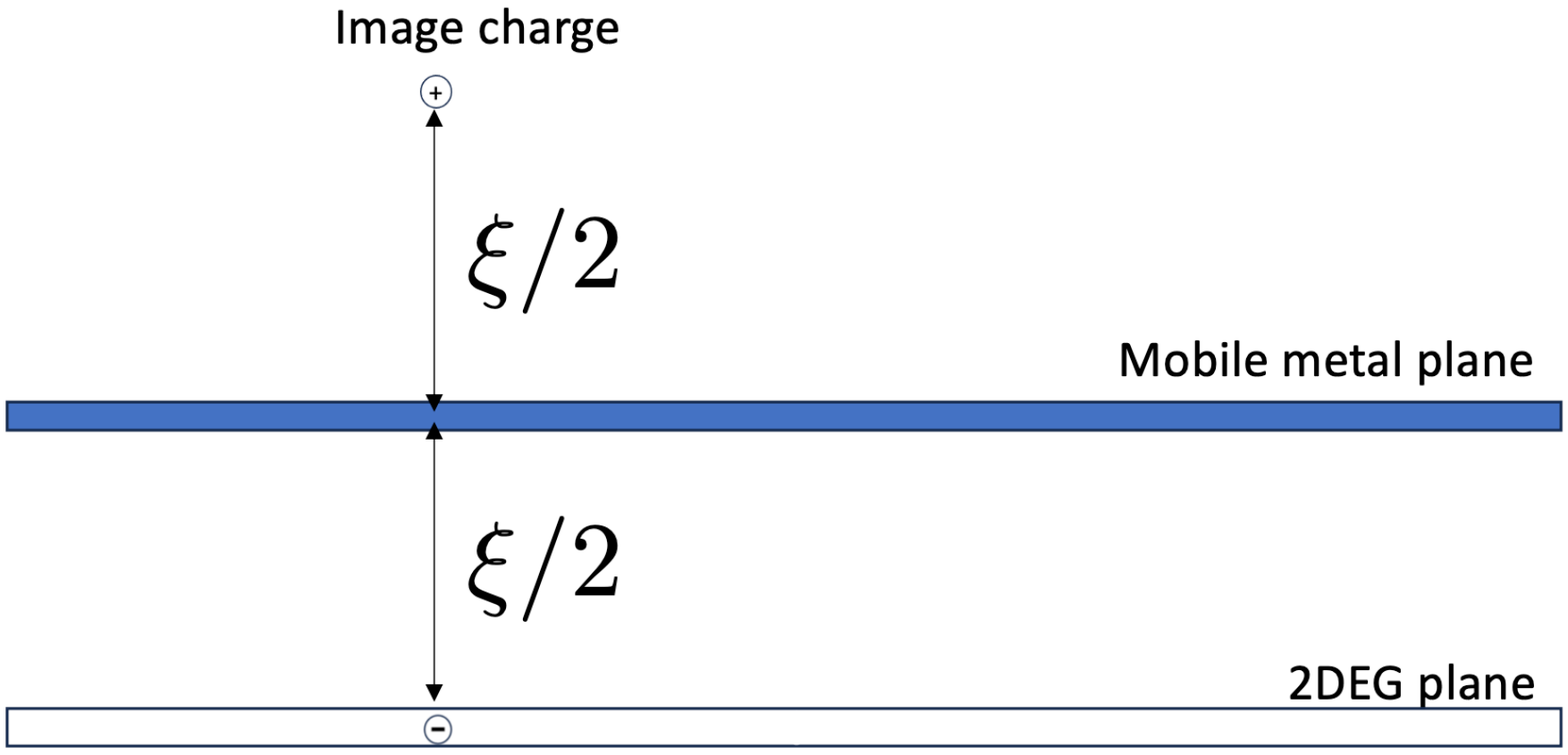}
\caption{Sketch depicting the image charge created by a mobile metal plate. $\xi$ is the distance between a charge in the 2D electron gas plane and its opposite image charge.   
}
\label{fig:sketch_image}
\end{figure}
Here, we present detailed calculations demonstrating that the electrostatic screening effects produced by our metallic hovering resonator are negligible. This is primarily due to the fact that the distance between the 2D electron gas and the mobile metal plane is significantly greater than several tens of times the cyclotron length at the relevant magnetic fields.

Let us calculate the electrostatic screening. A metal plane that is parallel and at a distance $\xi/2$ from the 2D electron gas screens the Coulomb potential of an electron due to an opposite image charge located at the same in-plane position but at an out-of-plane distance $\xi$ from the 2D electron gas (see the sketch in Fig \ref{fig:sketch_image}).
The screened electrostatic potential reads:
\begin{equation}
V_{\mathrm{scr}}(r; \xi) = \frac{e^2}{4 \pi \epsilon_0 \epsilon_r} \left [ \frac{1}{r} - \frac{1}{\sqrt{r^2 + \xi^2 }} \right ] \,.
\end{equation}
The related Haldane pseudopotentials on a disk geometry 
are:
\begin{equation}
    v_m = 
    \frac{1}{2^{2m+1} \Gamma(m+1)}
    \int_0^{+\infty}
    du\;
    u^{2m+1}
    V_{\mathrm{scr}}( u\ell; \xi )
    e^{-\frac{1}{4}u^2} \, .
\end{equation}

\begin{figure}[t!]
\centering
\includegraphics[width=0.5\textwidth]{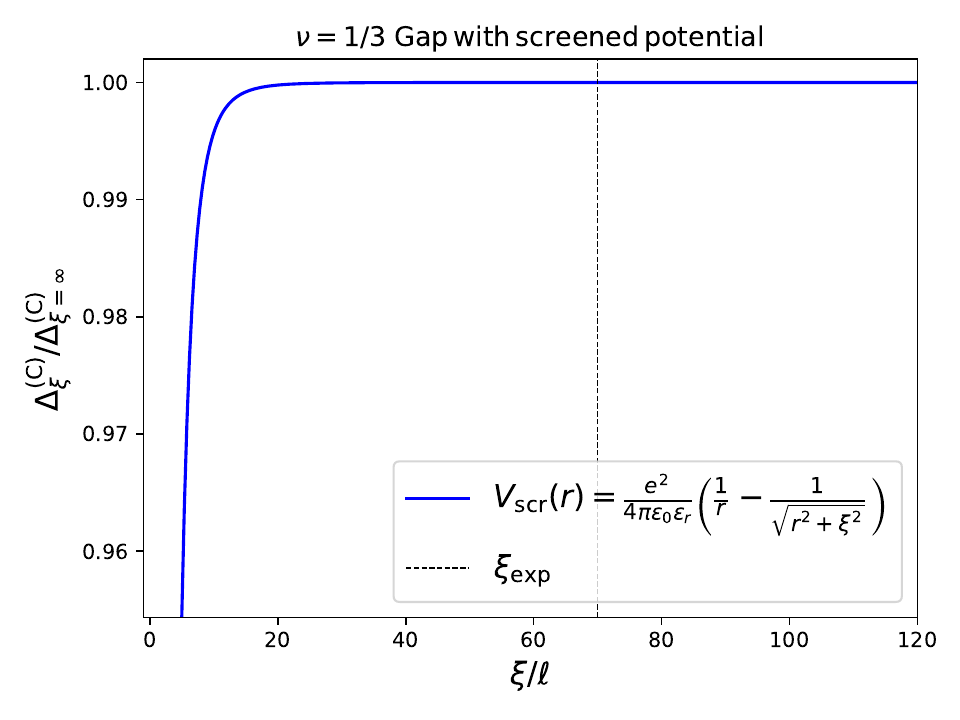}
\caption{
    Evolution of the 1/3 fractional quantum Hall energy gap with a Coulomb potential screened by a metal plane at a distance $\xi/2$. The gap is normalized to the same quantity calculated for the bare Coulomb potential. The distance is expressed in units of the magnetic length $\ell$. The result has been obtained by exact diagonalization with $N= 5$ electrons. 
    In the experiments $\xi >  2 d = 700$ nm (due to the GaAs buffer on top of the quantum well).
    For $B = 5$ T, the magnetic length $\ell \simeq 10$ nm. Hence, for our experimental set-up, we have $\xi/ \ell > 70$. As shown by the figure, the variation of the fractional quantum Hall gaps is totally negligible at such distances. Note also the tiny decrease of the gap at much shorter distances. 
}
\label{fig:screening}
\end{figure}

In our experimental configuration, the mobile metal plane is at a distance $\xi/2 > d \geq 350$ nm, 
$d$ being the distance from the 2DEG to the mobile metal plane,
where we have not even included the GaAs buffer ($250 \mathrm{nm}$-thick) separating the 2D electron gas from air. In the region of interest, the magnetic field is around $5$ T, yielding a magnetic length $\ell \approx 10$ nm. Hence, the ratio $\xi/\ell > 70$. At these distances, the electrostatic screening effects on the fractional quantum Hall phases should be completely negligible. The results of our exact diagonalization results in Fig. \ref{fig:screening} show exactly that. We report how the $1/3$ fractional quantum Hall gap changes as a function of the ratio $\xi/\ell$. In particular the gap with the electrostatically screeened potential is normalized to the gap with the bare Coulomb potential. The results show clearly that the variation of the gap for $\xi/\ell= 70$ is totally negligible. 
Note that moreover at short distances (order of the magnetic length) the fractional quantum Hall gap does not increase, but decreases, hence contrary to the experiments. 

Note also that our theoretical calculations agree with the results of a previous theoretical paper \cite{Abanin2011}, which studied the effect of a dielectric in close proximity to a 2D electron gas. They demonstrated that effects occur only when the distance is comparable to the magnetic length. Moreover, the metallic case corresponds to $\alpha = \frac{\epsilon_1-\epsilon_2}{\epsilon_1+\epsilon_2} = -1$ in Ref. \cite{Abanin2011} (for a metal the absolute value of the dielectric constant $\vert \epsilon_2 \vert \gg \vert \epsilon_1 \vert$ where $\epsilon_1$ is the dielectric constant of the 2D material). The results in Ref. \cite{Abanin2011} indeed show a reduction of the fractional quantum Hall effect rather than an increase for the case $\alpha = -1$.

% Interactions between electrons within the same Landau level 
% are fundamental to the phenomena observed in fractional quantum Hall effects. 
% Alterations in these effects are likely influenced by deviations 
% in the interaction potential from the standard Coulomb one.
% A possible cause of these changes is 
% the exposure of the two-dimensional electron gas (2DEG) 
% to an environment that either modulates or screens the potential, 
% thus impacting how the electrons interact \cite{Abanin2011}.
% One might think that the hovering cavity is responsible for
% the observed increased fractional gap.
% Here we w

\section{Exact diagonalisation result with an effective cavity-mediated potential}

\begin{figure*}[h!]
\centering
\includegraphics[width=1.0\textwidth]{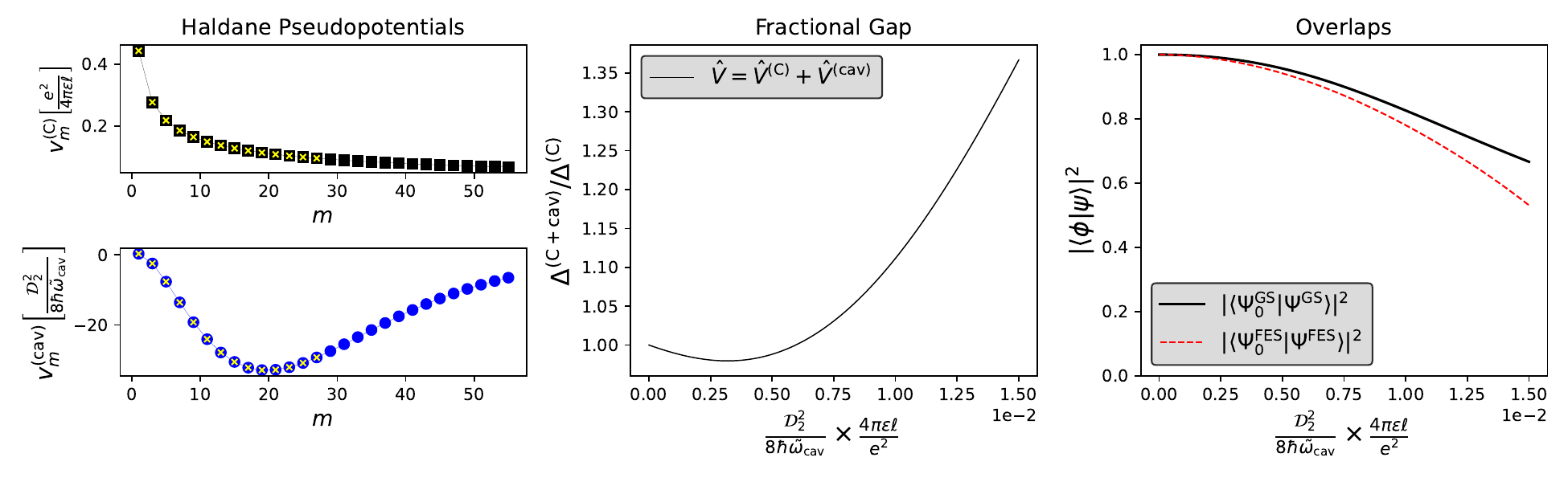}
\caption{
    Left panels: the odd Haldane pseudopotential components versus the angular momentum for the repulsive Coulomb interaction (top) and for the attractive cavity-mediated interaction (bottom) with an exponential gaussian cutoff (see text for details). The crosses indicate the angular momenta that count in the case with $N = 5$ electrons.
    Central panel: exact diagonalization results for the fractional quantum Hall gap with $N = 5$ electrons for filling factor equal to $1/3$. The gap is normalized to the gap without the cavity interaction. The horizontal axis displays the amplitude of the attractive potential. Right panel: the solid line depicts the overlap between the ground state with the cavity-mediated potential and the ground state with no cavity coupling; the red dashed line represents the overlap for the first excited state.
} 
\label{fig:three_plots}
\end{figure*}

Here we report exact diagonalization results for a small number of electrons 
in a disk geometry. We show that adding an attractive cavity-mediated potential to the Coulomb repulsive interaction we can get an enhancement of fractional quantum Hall gaps.
Let us consider our cavity-mediated potential
\begin{equation}
    \tilde{V}^{(\mathrm{cav})} ( r )
    =
    \left(
        -
        \frac{
            \mathcal{D}_2^2     
        }{
            8 \hbar \tilde{\omega}_{\mathrm{cav}}
        }
    \right)
    \left[
        \frac{1}{16}
        \left(
            \frac{r}{\ell}
        \right)^4
        - 
        \left(
            \frac{r}{\ell}
        \right)^2
        +
        2
    \right]
    e^{-\frac{1}{2} \left(\frac{r}{L_{\mathrm{c}}}\right)^2}
    \, ,
    \label{V_r}
\end{equation}
where we have introduced a gaussian cutoff characterized by the cut-off length $L_{\mathrm{c}}$ . Note that our potential has been calculated considering a constant spatial gradient of the vacuum electric field and would normally diverge when the distance $r$ goes to infinity. This is due to the simplifying approximation of a constant gradient. The gaussian cutoff regularizes such divergence. 

To perform exact diagonalization, it is convenient to get the Haldane pseudopotential components for such potential:
\begin{equation}
    \tilde{v}_m^{(\mathrm{cav})} 
    = 
    \frac{1}{2^{2 m + 1} \Gamma(m+1)} 
    \int_0^{+\infty} du \; u^{2 m + 1} \tilde{V}^{(\mathrm{cav})}(u \ell) e^{-\frac{1}{4} u^2}.
\end{equation}

After some algebra, the Haldane components read
\begin{widetext}
\begin{equation}
    \tilde{v}_m^{(\mathrm{cav})}
    =
    \left(
        -
        \frac{
            \mathcal{D}_2^2     
        }{
            8 \hbar \tilde{\omega}_{\mathrm{cav}}
        }
    \right)
    e^{ - \lambda_{\mathrm{c}} m }
    \left[
       e^{-3\lambda_{\mathrm{c}}} (m^2+3m+2)
       -4 e^{-2\lambda_{\mathrm{c}}}  (m+1)
       +
       2 e^{-\lambda_{\mathrm{c}}}
    \right] \, ,
\end{equation}
\end{widetext}
where $m_{\mathrm{c}} = (L_{\mathrm{c}}/\sqrt{2}\ell)^2$ and  
 $\lambda_{\mathrm{c}} = \log(\frac{m_{\mathrm{c}}+1}{m_{\mathrm{c}}})$. 
Note that the Gaussian cut-off for the potential in real space corresponds to an exponential cutoff of the Haldane components at large values of the angular momentum $m$.

Fig. \ref{fig:three_plots} reports exact diagonalization results for $5$ electrons. The left panels show the Haldane components for the Coulomb repulsive interaction (top) and the cavity-mediated attractive interaction (bottom). The central panel shows results for the $1/3$ fractional quantum Hall gap versus the amplitude of the cavity-mediated interaction. The gap is normalized to the value of the gap with no cavity coupling. The amplitude of the cavity-mediated potential is expressed in Coulomb units. Remarkably we do observe an enhancement of the gap. Note that for these calculations with only $5$ electrons, we use artificially large values of the cavity-mediated potential amplitude, because we cannot use exact diagonalization for the very large number of electrons that are present in the experiments. In particular, we cannot use it to explore collective effects due to the long-range potential.
The solid line in the right panel of Fig. \ref{fig:three_plots}  reports the overlap between the ground state modified by the cavity vacuum fields with the ground state without. It is apparent that there is a smooth and continuous decrease of such overlap. A gap increase by approximately $35 \%$  (central panel) corresponds to an overlap of about $70 \%$. The dashed line displays the overlap for the first excited state, which exhibits a similar behavior. 

\section{Magneto-roton theoretical results for the fractional quantum Hall gaps}

Here, we apply the magneto-roton theory of Girvin, MacDonald, and Platzman\cite{Girvin_master_piece} to estimate the variation in fractional quantum Hall gaps produced by the cavity vacuum fields. This theory is based on a successful ansatz, similar to the one introduced by Feynman to describe excitations in superfluid helium, but restricted to the lowest Landau level. The only inputs needed are $V(q)$, the Fourier transform of the electron-electron potential in momentum space, and $\overline{s}(k)$, the static structure function. The static structure function is related to the Fourier transform of the density-density correlation function (with the density operator projected onto the lowest Landau level) for the considered fractional quantum Hall ground state.

The Fourier transform of the real-space potential in Eq. (\ref{V_r}) reads: 
\begin{widetext}
\begin{equation}
    V^{\mathrm{(cav)}} ( q )
    =
    -
    \left(
        \frac{
            \mathcal{D}_2^2
        }{  
            8 \hbar \tilde{\omega}_{\mathrm{cav}}
        }
    \right)
    \left(
        2 \pi L^2
    \right)
    \left\{
        \frac{L^4}{16 \ell^4}
        (qL)^4
        +
        \left[
            -\frac{L^4}{2\ell^4}
            +
            \frac{L^2}{\ell^2}
        \right]
        (qL)^2
        +
        \left[
            \frac{L^4}{2\ell^4}
            -
            2 \frac{L^2}{\ell^2}
            +
            2
        \right]
    \right\}
    e^{-\frac{1}{2}(Lq)^2} \, .
\end{equation}
\end{widetext}

The static structure for the lowest Landau level reads

\begin{equation}
    \overline{s}( \bm{k} ) = s( \bm{k} ) - (1-e^{-\frac{1}{2}(\ell \vert \bm{k} \vert)^2})\, ,
\end{equation}
where 
\begin{equation}
    s( \bm{k} ) 
    =
    \rho (2 \pi)^2 \delta^{(2)}(\bm{k}) +
    1 - \rho \int d^2 r e^{-\img \bm{k} \cdot \bm{r}} [g(\bm{r})-1] \, 
    \end{equation}
with $\rho$ being the density of the 2D electron gas and $g(\bm{r})$ is the density-density correlation function of the ground state. 

With our long-range potential we need to regularize the theory by introducing a cutoff at long distances. For finite-size samples, we also have to regularize the Dirac delta distribution in the static function. For simplicity, we will consider the following represention of the delta function:

\begin{equation}
    \delta^{(2)}_L ( \bm{k} )
    =
    \frac{L^2}{\pi} \mathbf{1}_{\vert \bm{k} \vert < \frac{1}{L}} \, .
\end{equation}
Note that we have $\int d^2q \delta^{(2)}_L ( \bm{k} ) = 1$ 
and 
$\lim_{L\to + \infty} \delta^{(2)}_L ( \bm{k} ) = 0$ 
for all 
$\bm{k} \neq \bm{0}$.

The expression of the magnetoroton gap is:
\begin{equation}
\Delta = \min_k \frac{\overline{f}(k)}{\overline{s}(k)} \, ,
\end{equation}
where
\begin{align}
    \overline{f}(k)
    &=
    \nonumber
    \int
    \frac{d^2q}{(2 \pi)^2}
    V ( q )
    \left\{
        1
        -
        \cos
        \left[
            \ell^2 \left( \bm{k} \times \bm{q} \right)_z
        \right]
    \right\} \times \\
    &\times
    \left[
        \overline{s}( \bm{k} + \bm{q} )
        e^{\ell^2 \bm{k} \cdot \bm{q}}
        -
        \overline{s}( \bm{q} )
        e^{-\frac{1}{2}(\ell \bm{k})^2}
    \right] \, 
\end{align}
is the lowest Landau level projected oscillator strength.
Hence the variation of the gap due to the cavity is given by the expression:
\begin{equation}
\Delta^{(\mathrm{cav+C})} - \Delta^{(\mathrm{C})}  = \min_k \frac{\overline{f}^{(\mathrm{cav})}( k ) }{\bar{s}(k)}\, ,
\end{equation}
where in the limit $L \gg l$ we have
\begin{align}
    \nonumber
    \overline{f}^{(\mathrm{cav})}_L( k )
    &\simeq 
    \left(
        \frac{L}{\ell}
    \right)^4
    \left(
        \frac{
            \mathcal{D}_2^2
        }{
            8 \hbar \tilde{\omega}_{\mathrm{cav}}    
        }
    \right)
    (\ell k)^2 e^{-\frac{1}{2}(\ell k)^2} \times \\
    \nonumber
    &\times 
    \left(
        \frac{
            \nu
        }{
            8 \pi^2
        }
    \right)
    \left[
        \int_0^{1}
        d \eta \;
        \left(
            \frac{1}{16} \eta^7 - \frac{1}{2} \eta^5 + \frac{1}{2} \eta^3
        \right)
        e^{-\frac{1}{2} \eta^2 } 
    \right] \\
    &\simeq
    0.04
    \left(
        \frac{L}{\ell}
    \right)^4
    \left(
        \frac{
            \mathcal{D}_2^2
        }{
            8 \hbar \tilde{\omega}_{\mathrm{cav}}    
        }
    \right)
    (\ell k)^2 e^{-\frac{1}{2}(\ell k)^2} 
    \left(
        \frac{
            \nu
        }{
            8 \pi^2
        }
    \right)
    \, .
\end{align}

\begin{figure}
    \centering
    \includegraphics[scale = 0.5]{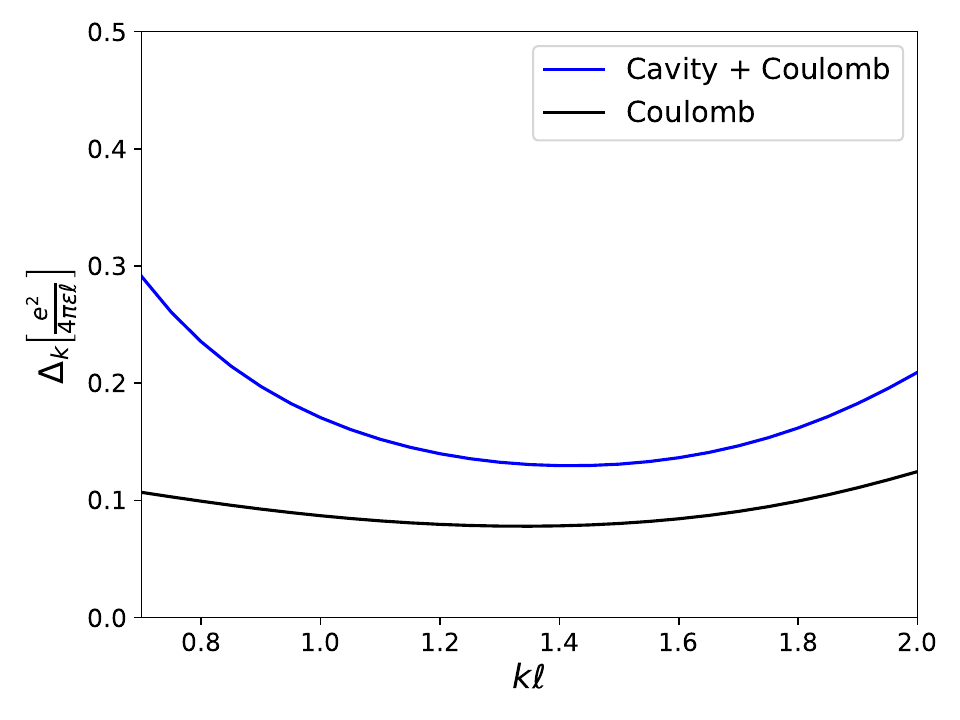}
    \caption{Magnetoroton energy-momentum  dispersion for the $1/3$-state without cavity (black line) and with the cavity (blue line). Parameters are the same discussed in the main text. }
    \label{fig:magnetoRotonGap}
\end{figure}

We point out that that the long-range nature of the cavity-mediated interaction is responsible for the extensive scaling of the variation of the excitation gap. Note that the $k$-dependence of  $\overline{f}^{(\mathrm{cav})}_L( k )$ is smooth and does not shift significantly the magneto-roton minimum wavevector $k_{\mathrm{min}}$ even when the gap is increased by $50 \%$ (see Fig. \ref{fig:magnetoRotonGap}).

A more detailed derivation of the theory with additional studies will appear in a forthcoming publication \cite{borici_in_preparation}.

\bibliography{scibib}% common bib file
\bibliographystyle{apsrev4-1}